\documentclass{aa}
\usepackage[colorlinks=true, linkcolor=blue, citecolor=blue, filecolor=blue,urlcolor=blue]{hyperref}
\usepackage{graphicx,natbib,txfonts,color,xspace,amsmath,mathtools,afterpage,booktabs,multirow,caption,url}

\bibpunct{(}{)}{;}{a}{}{,}

\begin{document}
        \title{Hidden oscillations in plain sight: identification of seismically unresolved red-giant asteroseismic binary candidates}
        \author{Jeong Yun Choi\inst{1,2}, Francisca Espinoza-Rojas\inst{1,2} \and Saskia Hekker\inst{1,2}}
        \institute{Heidelberg Institute for Theoretical Studies (HITS), Schloss-Wolfsbrunnenweg 35, D-69118 Heidelberg, Germany\\ 
        \email{jeongyun.choi@h-its.org} 
        \and 
        Heidelberg University, Centre for Astronomy, Landessternwarte, K\"onigstuhl 12, D-69117 Heidelberg, Germany} 
        \authorrunning{J.Y. Choi et al.}
        \titlerunning{Seismically unresolved red-giant asteroseismic binaries}

        \abstract{Light curves of oscillating stars provide valuable insights into the stellar interiors. In some cases, the aperture mask used to extract a light curve can record fluxes from multiple stars. When oscillations from a pair of stars are captured within a single aperture, they can be considered as potential asteroseismic binaries (ABs). If the two stars oscillate at similar frequency ranges, the superpositioned oscillation patterns appear as if from a single star, which can lead to the derivation of inaccurate asteroseismic parameters. In this study, we investigate seismically unresolved AB candidates consisting of two red-giant stars observed by \textit{Kepler}. We aim to make a direct comparison between the power density spectra (PDSs) of blended and separated oscillations from both stars in each candidate, and examine how oscillations from two stars in a PDS impact the measurement of asteroseismic and stellar parameters. We selected stars from the APOKASC3 catalog that have at least one neighboring source within 20 arcsec and show oscillations in similar frequency ranges. We focus on the systems where the light curves from each star in AB candidates are available or can be extracted with a custom mask. From these star pairs, we identified 6 seismically unresolved AB candidates whose PDS morphologies change noticeably across light curves extracted with different apertures. We determined asteroseismic parameters and derived masses and radii from the PDSs of both AB candidates and single stars to quantify the biases. Oscillations from two stars in a single PDS cause inaccurate mode identification and bias the seismic parameters. These biases propagate into stellar properties, with mass and radii for the 6 AB candidates investigated here differing by up to $\sim$3 and $\sim$2 times relative to the individual stars, respectively. For the AB candidate with the most complex PDS, core properties such as period spacings and coupling factors become unreliable, with the coupling factor often being overestimated. We checked that all 6 AB candidates are chance alignments. Our results indicate that the inconsistencies in asteroseismic and stellar parameters across various studies on the stars with complex PDSs can be explained by potential seismically unresolved asteroseismic binaries. These findings highlight the importance of identifying and accurately accounting for such systems in asteroseismic analysis.}
        \maketitle

\section{Introduction}
Evolved cool red-giant stars have deep convection zones in their outer layers and exhibit stellar oscillations. Like the Sun, these oscillations are stochastically driven by near-surface convection and are called solar-like oscillations (\citealt{2019_Garcia_Ballot}). The combination of waves that are refracted by the internal density gradients and reflected at the surface forms standing waves at discrete frequencies. By observing the brightness or radial velocity variations, we can detect these wave patterns. From NASA’s \textit{Kepler} mission \citep{2010_Borucki}, about 21,000 oscillating red giants have been reported (\citealt{2011_Hekker, 2011_Huber, 2014_Huber, 2013_Stello, 2014_Pinsonneault, 2018_Pinsonneault, 2025_Pinsonneault, 2016_Mathur, 2016_Yu, 2018_Yu, 2019_Hon}). Moreover, \citet{2025_Vrard} recently analyzed the evolutionary stages of the red-giant \textit{Kepler} legacy catalog (Garcia et al., in prep.) which contains 30,337 red-giant candidates. This large number of oscillating red giants from \textit{Kepler} provide the opportunity to study and characterize the internal structures of these stars (see reviews by \citealt{2013_Chaplin_Miglio}, and \citealt{2017_Hekker}).

Given the 4 arcsec pixel size of \textit{Kepler} and the consequence of using a large photometric aperture for the targets, the extracted light curves often include flux from sources other than the target star that fall within the aperture. For example, \citet{2013_Gaulme} identified false positives in which red giants that appeared to belong to eclipsing binaries were in fact unrelated contaminants, by checking the contamination in the \textit{Kepler} target pixel files (TPFs\textcolor{black}{\footnote{\textcolor{black}{Target pixel files are stacks of exposures that contain pixel cutouts centered on a single star.}}}). Similarly, \citet{2016_Abdul-Masih} found stars whose light curves mimic eclipsing binaries and it turned out they are contaminated by nearby eclipsing binary sources. Moreover, \citet{2017_Colman} showed that anomalous high amplitude peaks detected in 168 red giants originate from other stars or possibly from stellar companions in close binary systems. In the study by \citet{2019_Hon}, oscillations detected in 909 dwarf/subgiant stars were categorized as blended targets which resembled red-giant oscillations. These oscillations were found to originate from nearby red giants rather than the main targets. More recently, \citet{2025_Wang} reported that the transit-like signal from KOI-1755 was a false positive caused by an eclipsing binary. Likewise, precise extraction of light curves is essential in a densely populated field, which requires techniques such as custom aperture photometry (e.g., \citealt{2022_Colman, 2023_Covelo-Paz}) or Point Spread Function (PSF) photometry (e.g., \citealt{2021_Hedges}). 

Recently, \citet{2023_Martinez_Palomera} provided KBonus-Background\textcolor{black}{\footnote{\textcolor{black}{Throughout this paper, we used "KBonus" as shorthand for "KBonus-Background".}}} light curves, which are extracted by using the Linearized Field De-blending method of \citet{2021_Hedges} as implemented in the Python package \texttt{psfmachine} \citep{2021_Hedges_psfmachine}. KBonus offers various opportunities, such as mitigating dilution which yields deeper transit depth of a planet or deblending contamination which is useful to distinguish false positives from real exoplanet candidates. Additionally, \citet{2025_Borkovits} analyzed eclipse timing variation curves of \textit{Kepler} eclipsing binaries and extended the available \textit{Kepler} data to span a larger fraction of the full mission by using KBonus light curves for a subset of stars. Moreover, \citet{2025_Claytor} used KBonus light curves to estimate rotation periods of new background sources. The presence of such a large number of background sources highlights the importance of accurately defining the photometric aperture when measuring the flux of a star. 

When the photometric aperture contains more than one oscillating star, the extracted light curve records oscillation signals from all enclosed stars. Consequently, their oscillation signatures are all present in the power density spectrum (PDS) of the target star. This occurs when a pair of stars happen to lie at a small angular separation either by chance (i.e., chance alignment), or because the stars are gravitationally bound (\citealt{2015_Appourchaux, 2017_White, 2018_Li, 2018_Marcadon, 2016_Rawls, 2018_Beck, 2018_Themessl_2568888, 2025_Grossmann, 2025_Espinoza-Rojas, 2026_Schimak}). We consider such cases as asteroseismic binary (AB) candidates in which a single light curve contains oscillations from two stars. 

Population synthesis studies of the \textit{Kepler} field by \citet{2014_Miglio} predicted at least 200 ABs in the \textit{Kepler} long-cadence data. Most of these predicted ABs consist of red-giant stars showing detectable oscillations, and the fraction of physically unassociated pairs is expected to be low (less than 10\% of detectable ABs; \citealt{2014_Miglio}). However, \citet{2025_Espinoza-Rojas} analyzed 40 red-giant AB candidates that show two distinct power excesses in their PDSs and found only two wide-binary candidates, while most cases are chance alignments. 

Following \citet {2014_Miglio}, \citet{2025_Mazzi} concluded that ABs with two core-He-burning (CHeB) have the largest fraction of their simulation as the two components of a gravitationally bound system have similar masses, radius, and luminosity. In particular, red-clump (RC) stars have similar core masses and luminosities after the helium flash, so they exhibit oscillations in similar frequency ranges. Therefore, we expect to observe seismically unresolved\footnote{We define “seismically unresolved” as PDS in which the detectable oscillation signatures of two stars overlap and show a single power excess.} red-giant ABs when two such stars fall within a single photometric aperture. 

Seismically unresolved ABs can be easily mistaken for single stars due to the combined oscillation signals. Even when they are identified, the overlapped oscillation patterns can bias the measurement of the frequency of maximum oscillation power, $\mathrm{\nu_{max}}$ \citep{2019_Sekaran}. In addition, seismically unresolved red-giant pairs with similar $\mathrm{\nu_{max}}$ and brightness tend to show complex PDS morphologies \citep{2025_Choi}. The lack of clear structures in the PDS can further bias the measurement of the frequency separation between modes of the same spherical degree and consecutive radial orders ($\mathrm{\Delta\nu}$). Since asteroseismic scaling relations \citep{1991_Brown,1995_Kjeldsen} depend on $\mathrm{\nu_{max}}$ and $\mathrm{\Delta\nu}$, biases in these parameters directly propagate into the stellar parameters. 

Furthermore, blended oscillation signals from two stars complicate the mode identification. Red-giant stars exhibit mixed-modes which behave as acoustic pressure modes ($p$-modes) in the envelope and as gravity modes ($g$-modes) in the core. The $g$-dominated mixed modes have larger amplitudes in the core than in the envelope. Similar to pure $g$-modes, these modes are almost equally spaced in period and their average period spacing is useful to probe red-giant cores \citep{2011_Beck, 2011_Bedding, 2014_Mosser}. Consequently, the inferred core properties can be systematically biased if we do not take potential ABs into account. 

In this paper, we identify seismically unresolved red-giant asteroseismic binary candidates from \textit{Kepler} data and quantify the biases that arise when these are not regarded as AB candidates. We focus on the candidates where we can obtain (i) light curves extracted from an aperture that contains flux from both stars, so that the PDSs represent the combined oscillation signals of the AB candidates, and (ii) light curves in which only the flux of a single star contributes, so that their PDSs show the oscillations of the individual stars (Sect. \ref{sec:Data}). This allows us to compare the PDSs of individual stars to the PDSs of AB candidates and investigate how asteroseismic parameters (determined in Sect. \ref{sec:method}) vary due to the influence of the overlapped oscillations (Sect. \ref{sec:Results}). We assess whether the stars in the identified seismically unresolved AB candidates are gravitationally bound in Sect. \ref{sec:Discussions}, and conclude in Sect. \ref{sec:Conclusion}.

\section{Data preparation and identification of AB candidates} \label{sec:Data}
Here, we first describe the target selection from the \textit{Kepler} long-cadence data. We subsequently explain the preparation of the light curves of the individual stars and of the AB candidates as well as the methods to compare morphologies of the PDSs. Finally, we present the identification of seismically unresolved red-giant AB candidates and examine \textit{Gaia} DR3 multiplicity indicators \citep{2023_Gaia_multiplicity} to assess the gravitationally bound nature of each system.

\subsection{Target selection}\label{subsec:candidate_AB}
We investigated stars listed in the APOKASC3 \citep{2025_Pinsonneault} catalog. APOKASC3 combined \textit{Kepler} asteroseismology with APOGEE spectroscopy (DR 16 and 17 \citealt{2020_Ahumada, 2022_Abdurrouf}) to derive robust stellar parameters. Around the \textit{Kepler} coordinates of each APOKASC3 target, we searched for other APOKASC3 stars located within a radius of 20 arcsec (equivalent to $\sim$ 5 \textit{Kepler} pixels) with 10 < \textit{Kepler} magnitude (Kp) $\leq$ 17. This magnitude range avoids saturated pixels and ensures the detectability of the oscillations, considering that faint red giants can have Kp as low as $\sim$16 mag \citep{2016_Mathur}.
 
To find the AB candidates with overlapping oscillations, we compared the $\mathrm{\nu_{max}}$ values reported in APOKASC3 catalog for each target star and its selected nearby stars. When the oscillation power excess is approximated as a Gaussian envelope (see Sect. \ref{subsec:background}), overlap begins when the separation in $\mathrm{\nu_{max}}$ between the two stars is smaller than half the sum of the two envelope widths. For \textit{Kepler} red giants, the width of the power excess is known to scale as a power law of $\mathrm{\nu_{max}}$ (\citealt{2012_Mosser}; width $\mathrm{\approx 0.66\nu_{max}^{0.88}}$). This implies that the overlap can occur for star pairs with $\mathrm{\nu_{max}}$ differences of up to $\sim40\%$. However, KIC 7345204, identified by \citet{2025_Espinoza-Rojas} as a seismically resolved AB candidate, shows two very close distinct power excesses with a $\sim40\%$ difference in $\mathrm{\nu_{max}}$. Therefore, we adopted conservative threshold and selected pairs of stars where the difference in $\mathrm{\nu_{max}}$ between the two stars is less than 30\% of their mean $\mathrm{\nu_{max}}$. This allows us to focus on seismically unresolved AB candidates where their power excess can be modelled as a single Gaussian envelope, while allowing the possibility that the reported $\mathrm{\nu_{max}}$ values may be biased due to the oscillations from two stars. As a result, we selected 40 seismically unresolved AB candidates.

\subsection{Light curves preparation} \label{subsec:Data-lightcurve}
We collected \textit{Kepler} long-cadence light curves from MAST (Mikulski Archive for Space Telescopes) extracted by different pipelines: KEPSEISMIC\footnote{{https://archive.stsci.edu/prepds/kepseismic/}}, KBonus, and PDCSAP (Presearch data conditioning simple aperture photometry; \citealt{2012_Stumpe}). KEPSEISMIC light curves were corrected using the \textit{Kepler} Asteroseismic data analysis and calibration Software (KADACS; \citealt{2011_Garcia}). The correction involves removing outliers, addressing jumps and drifts, and stitching data from all quarters. It also fills various sizes of the regular gaps in long cadence data using a multiscale discrete cosine transform (\citealt{2014_Garcia}), which arise from \textit{Kepler} nominal mission operations. We used the light curves filtered using a 20-day high-pass filter. 

\begin{figure*}[!htbp]
    \centering
    \vspace*{-0.02\linewidth}
    \includegraphics[width=0.97\linewidth]{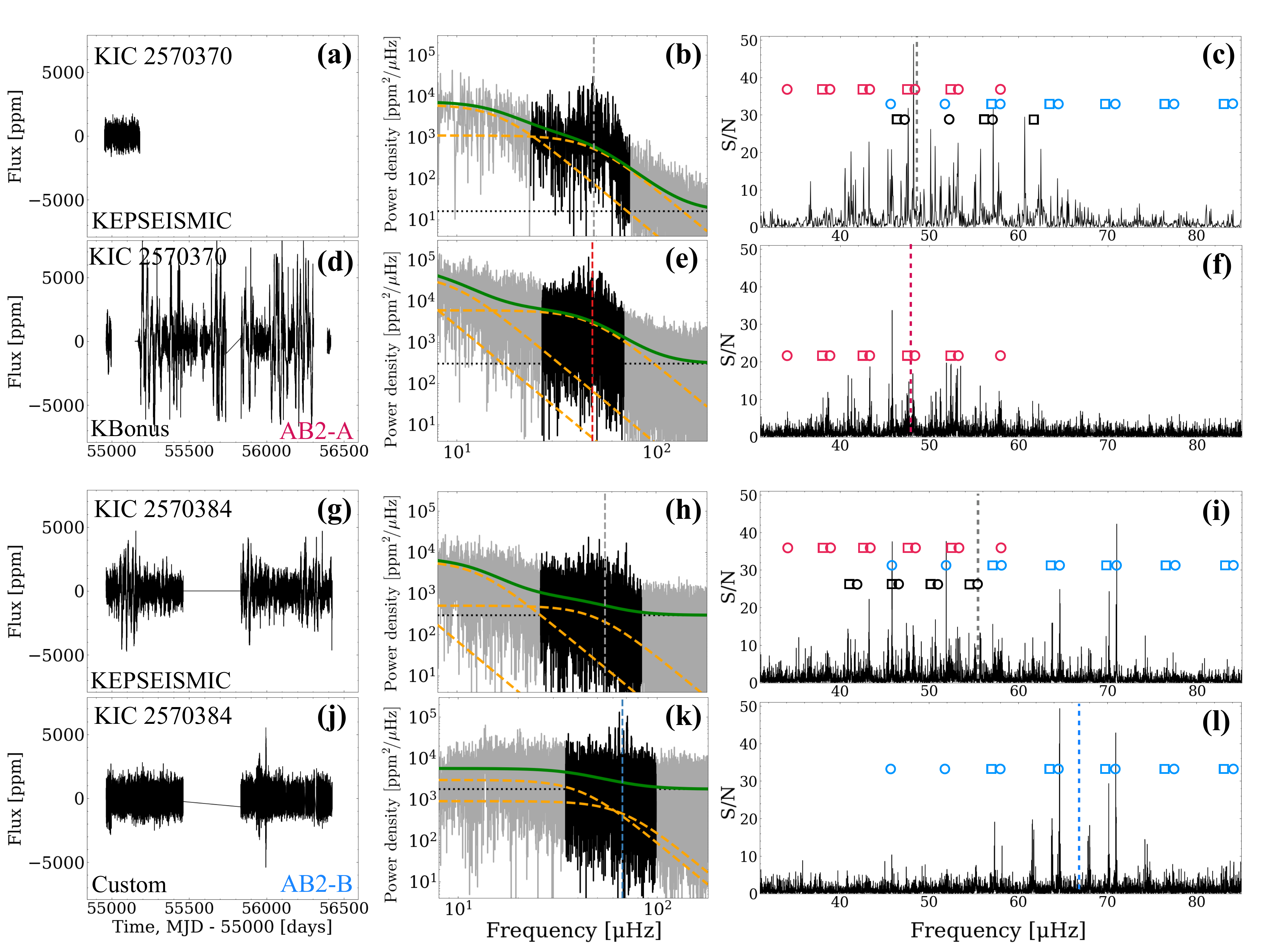}
    \caption{Seismically unresolved AB candidate consisting of KIC 2570370 (AB2-A) and KIC 2570384 (AB2-B). Panels (a), (d), (g) and (j): Light curves of AB2 ((a) and (g)), AB2-A (d) and AB2-B (j). Panels (b), (e), (h) and (k): PDS of each light curve. The total background fit (green solid lines) is composed of granulation background components (orange dashed lines), and white noise (black dotted lines). The frequency ranges of $\mathrm{\nu_{max} \pm 3\Delta\nu}$ are highlighted in black. Panels (c), (f), (i), and (l): Background-normalized PDSs of the highlighted frequency range in panels (b), (e), (h) and (k), respectively. The modes are marked by circles ($\mathrm{\ell=0}$) and squares ($\mathrm{\ell=2}$). Those peaks identified in the PDSs from KEPSEISMIC light curves are indicated in black, while those of individual stars are in red (AB2-A) and blue (AB2-B). The vertical dashed lines indicate the position of $\mathrm{\nu_{max}}$.}
    \label{fig:2570370_2570384_PDS}
\end{figure*}

To check whether the KEPSEISMIC light curves contain oscillation signals from both stars, we first used KBonus light curves for comparison. When multiple oscillating stars are closely located on the TPF, KBonus light curves are particularly useful because they are specifically designed to de-blend the contributions from multiple sources. Furthermore, they often provide longer timespan of the data for the targets, since they use all neighboring TPFs that contain fluxes of the targets (see the difference in Figs. \ref{fig:2570370_2570384_PDS}(a) and \ref{fig:2570370_2570384_PDS}(d)). \citet{2023_Martinez_Palomera} used \textit{Gaia} DR3 as the input sources to extract KBonus light curves. They corrected velocity aberration and modelled the time-dependent perturbation of the instrumental PSF for each \textit{Gaia} DR3 source, following the method from \citet{2021_Hedges}. For our analysis, we used the corrected PSF light curves\footnote{\textcolor{black}{{KBonus light curves are available from MAST as a High Level Science Product (HLSP) with DOI: 10.17909/7jbr-w430}}}, which are stitched across multiple quarters. We filtered out flagged cadences with any quality flags and removed 5$\sigma$ flux outliers. 

Although KBonus light curves are designed to provide de-blended signals, the light curves can still be contaminated by neighboring sources, particularly in crowded fields. Moreover, KBonus light curves are less reliable for bright stars (Kp < 12 mag), or when the fraction of the PSF on the TPF is low (\texttt{psffrac} < 0.6). In such cases, we checked other available light curves (i.e., PDCSAP light curves provided by the \textit{Kepler} pipeline; \citealt{2010_Jenkins}) to see whether their PDSs show oscillation signals from a single star. The PDCSAP light curves are corrected by using cotrending basis vectors \citep{2012_Smith} to remove systematic trends from the simple aperture photometry light curves. If the PDSs of the PDCSAP light curves also show oscillation signals from two stars, we extracted new light curves from the TPFs by creating custom apertures to construct the PDSs of individual stars in AB candidates. 

Our pipeline to create custom\textcolor{black}{\footnote{\textcolor{black}{We refer the extracted light curve by creating custom apertures as "custom" throughout this paper.}}} apertures begins by combining the TPFs of two stars using a modified version of \texttt{MOSAIC} pipeline, which was originally developed for K2 data \citep{2020_k2mosaic}. We input the TPFs which share the same channel and cadence to produce a single combined file. This allows us to define two non-overlapping aperture masks on a single TPF and to extract the fluxes of two stars simultaneously. We locate each star by transforming its \textit{Kepler} RA and DEC to the pixel coordinates of the combined TPF. We re-gridded the combined TPF onto a finer grid and set the center of the apertures on the sub-pixel nearest to the position of the target star. For each star, we empirically defined a slightly smaller than 3$\times$3 \textit{Kepler} pixel size aperture. When two apertures overlap on the combined TPF, we shifted each aperture in opposite directions. Subsequently, we extract the light curve for individual stars by summing the fluxes within each aperture. For quarters where only one star of the AB candidate has an available TPF, we skip the combining step and apply the same procedure to the single TPF. For more details on the correction applied to the light curves as well as the stitching of different quarters, we refer the reader to Appendix \ref{sec:appendix-TPFs}. 

\begin{figure*}[!htbp]
    \centering
    \includegraphics[width=0.96\linewidth]{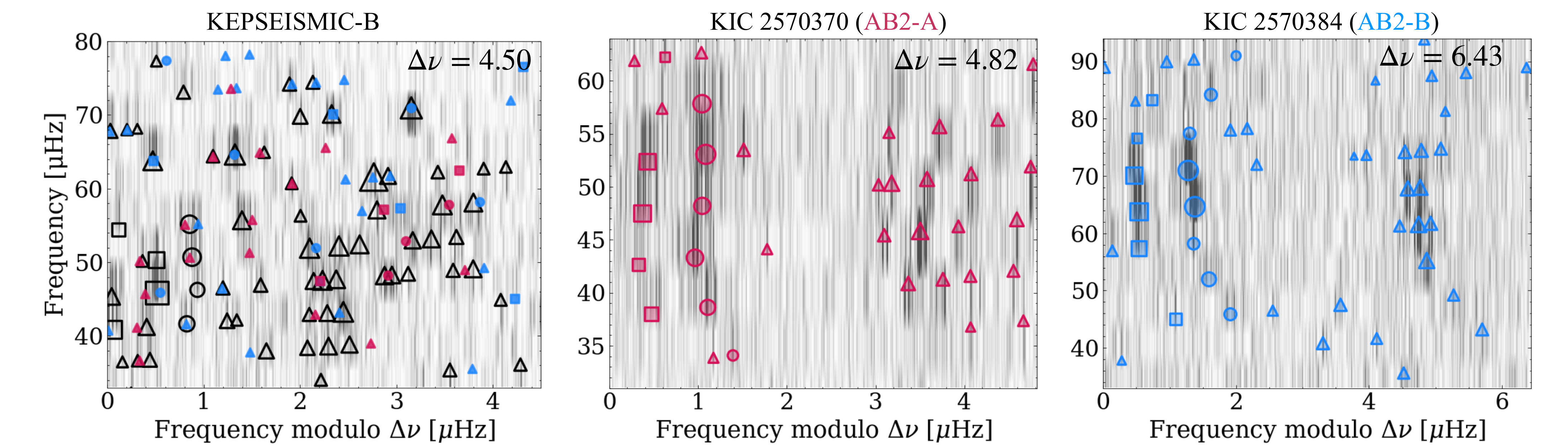}
    \caption{Frequency {\'e}chelle diagrams of AB2 (left), KIC 2570370 (AB2-A, middle) and KIC 2570384 (AB2-B, right). The background intensity in these {\'e}chelle diagrams is a smoothed representation of the PDSs, computed using the \texttt{echelle} Python package \citep{daniel_hey_echelle}. Each panel used different $\mathrm{\Delta\nu}$ values (see Sect. \ref{subsec:modeID}). Identified modes of the seismically unresolved AB are marked with black, and the modes for AB2-A (red) and AB2-B (blue) are overplotted in the left panel. The symbol sizes are scaled by the mode amplitudes, with $\ell=0, 1,$ and 2 modes plotted as circles, triangles, and squares, respectively.}
    \label{fig:2570370_2570384_frequency_echelle}
\end{figure*}

\subsection{Identification of seismically unresolved red-giant AB candidates} \label{subsec:identify_AB}
When oscillation signals from two stars overlap in a single PDS, we expect their PDS morphology to differ from those of the individual stars. Therefore, we compared the PDS morphologies from different light curves as described in Sect. \ref{subsec:Data-lightcurve} for the stars in the 40 AB candidates selected in Sect. \ref{subsec:candidate_AB}. We first excluded one AB candidate because neither star has a light curve that spans at least 10 quarters in any of the KEPSEISMIC, KBonus, and PDCSAP light curves. Then, we found 17 other AB candidates where the reported $\mathrm{\nu_{max}}$ values for two different KIC IDs are similar and their PDSs derived from KEPSEISMIC light curves show almost identical oscillation patterns. In each of these 17 cases, one of the two stars is flagged as a background source in APOKASC3, based on the $\mathrm{\nu_{max}}$ value that is inconsistent with the spectroscopic measurements. Additionally, we identified another AB candidate consisting of KIC 10096243 and KIC 10096249, where the two stars have similar reported $\mathrm{\nu_{max}}$ values. However, KIC 10096243 is also flagged as a background source in APOKASC3. We extracted custom light curves of KIC 10096243 and KIC 10096249 and confirmed that only KIC 10096249 showed a power excess with $\mathrm{\nu_{max}}$  $\sim36\mu$Hz. We excluded these total 18 candidates because the PDSs of the background sources show one oscillation power excess that actually originates from the red-giant star located nearby in the same TPF (see detail in Appendix \ref{appendix:false-positive}).

One AB candidate showed two distinctive power excesses in the PDS: KIC 9893437 and KIC 9893440. Due to the similar $\mathrm{\nu_{max}}$ values reported in APOKASC3 ($232.9\pm7.9\text{ }\mathrm{\mu Hz}$ for KIC 9893437, and $230.4\pm7.2\text{ }\mathrm{\mu Hz}$ for KIC 9893440), these stars were included as an AB candidate. In fact, the intrinsic $\mathrm{\nu_{max}}$ values of KIC 9893437 and KIC 9893440 are $232.9\pm0.9\text{ }\mathrm{\mu Hz}$ and $33.3\pm0.1\text{ }\mathrm{\mu Hz}$, respectively, according to \citet{2025_Espinoza-Rojas}.

We found 7 stars reported in the cluster member catalog of \citet{2020_Cantat-Gaudin}: KIC 2569891, 2569945, 2570384 in NGC 6791, and KIC 5024582, 5112467, 5112491, and 5112734 in NGC 6819. The KBonus light curves of these stars have low CROWDSAP values (< 0.4), which represents the fraction of target flux compared to the total flux within the photometric aperture including contaminating sources \citep{2023_Martinez_Palomera}. This indicates severe contamination from many nearby stars in densely populated fields. Attributing the observed oscillations to a specific star is beyond the scope of this work. Therefore, we excluded these stars from our main analysis except KIC 2570384. For KIC 2570384, the TPF contains only one additional star (KIC 2570370), which allows us to analyze these stars in more detail.

For the remaining 17 AB candidates, we compared the estimated values of $\mathrm{\nu_{max}}$ and $\mathrm{\Delta\nu}$ between the PDSs derived from KEPSEISMIC, KBonus, PDCSAP and custom light curves (see Sect. \ref{sec:method}  for the estimation of these parameters). When these values and the PDS morphologies across different light curves are consistent, we interpret that the observed oscillation peaks originate only from the target star itself and are not contaminated by another source. We flagged stars which have the percentage differences larger than 4\% in either $\nu_\mathrm{max}$ or $\Delta\nu$. For these cases, we visually inspected the other stars in each candidate to check whether the identified $\ell=0$ and 2 mode peaks of one star appear at the corresponding frequencies in the PDS of the other star. Finally, we found that the PDSs of the 6 AB candidates based on the KEPSEISMIC light curves clearly show overlapping oscillation signals from two red giants (see Table. \ref{tab:obs_info}). 

For example, the PDS of KIC 2570384 shows significantly different oscillation patterns between the PDSs of the KEPSEISMIC light curve and the custom light curve (see Fig. \ref{fig:2570370_2570384_PDS}(i) and (l)). This discrepancy arises because the KEPSEISMIC light curve of KIC 2570384 is contaminated by the flux from KIC 2570370, and thus the PDS of the KEPSEISMIC light curve shown in Fig. \ref{fig:2570370_2570384_PDS}(i) shows overlapping oscillation signals from both stars. Since all KEPSEISMIC light curves of the 6 identified AB candidates (listed in Table. \ref{tab:obs_info}) exhibit oscillations from two stars, we used the PDSs of KEPSEISMIC light curves as the PDSs of seismically unresolved AB cadidates throughout the paper. For individual stars, we used PDSs from different light curves (indicated in Table. \ref{tab:seismic_analysis} next to the relevant star IDs) favoring the PDS that showed the least amount of blended oscillation signals as described in Sect. \ref{subsec:Data-lightcurve}. For convenience, we label the six AB candidates from AB1 to AB6 and denote the two stars in each AB candidate as ABi-A for the brighter star and ABi-B for the fainter star (i = 1-6; see Table. \ref{tab:obs_info}).

\subsection{Flagged binary candidates in \textit{Gaia} DR3} \label{subsec:Gaia_check}

We queried the \texttt{gaiadr3.gaia\_source} table by using \textit{Gaia} DR3 IDs of the 6 seismically unresolved AB candidates to check the RUWE (Renormalized Unit Weight Error) values and the \texttt{non\_single\_star} flag \citep{2023_Gaia_multiplicity}. RUWE quantifies how well the \textit{Gaia} single-star astrometric solution fits the data, and therefore helps to flag binary or non-single star candidates. Notably, KIC 2449558 (AB1-A), KIC 3122188, and Gaia DR3 2051928734562919168 (AB3-B) have RUWE values larger than 1.4 (see Table. \ref{tab:obs_info}), indicating potential spatially unresolved binary systems \citep{2021_Lindegren}. 

We note that AB1-A (RUWE $\sim$1.8) is flagged as a non-single star system in the \textit{Gaia} DR3 astrometric acceleration catalog with an astrometric acceleration solution (acceleration model with 7 parameters; \citealt{2023_Halbwachs}). This indicates the presence of an unresolved gravitationally bound companion with an orbital period exceeding the \textit{Gaia} mission time interval. Importantly, this does not imply that AB1-A is gravitationally bound to AB1-B. We consider AB1-A as a star that has its own unresolved companion. Nevertheless, we retain AB1 (AB1-A and AB1-B) as an AB candidate throughout the paper.

For AB3 which consists of KIC 3122185 (AB3-A), we found two additional stars within the TPF. One is KIC 3122188, and the other is identified by the \textit{Gaia} ID 2051928738864291072. The corrected \textit{Gaia} BP and RP (Blue and Red Photometers) flux excess factor of KIC 3122188 is higher than the typical scatter level (\citealt{2021_Riello}), which indicates the presence of background contamination in the photometry. Due to their small angular distance ($\approx1$ arcsec), it is challenging to disentangle the flux contributions of KIC 3122188 and \textit{Gaia} ID 2051928738864291072. Based on the spectroscopic parameters of these two stars in Table. \ref{tab:obs_info}, the estimated $\mathrm{\nu_{max}}$ of KIC 3122188 is expected to be higher than that of \textit{Gaia} ID 2051928738864291072 and may lie above the Nyquist frequency ($\mathrm{\simeq283\mu Hz}$) of the \textit{Kepler} long-cadence data. Therefore, we adopt KIC 3122185 (AB3-A) and \textit{Gaia} ID 2051928738864291072 (AB3-B) in a pair for the rest of the paper, rather than pairing KIC 3122185 (AB3-A) with KIC 3122188 (see also Appendix \ref{appendix:AB3} and Fig. \ref{fig:3122188_TPF}). 

\begin{table*}[htbp]
\centering
\caption{Observational properties of 6 seismically unresolved ABs}
\resizebox{0.99\linewidth}{!}{
\begin{tabular}{c c c c c c c c c c}
\hline\hline
KIC &  Name & \textit{Gaia} DR3 ID &  Kp & \textit{G}mag & $T_\mathrm{eff}$ [K] &  log $g$ [dex] & [Fe/H] & RUWE & $s$ [AU] \\  
 \midrule
2449558 &   AB1-A & 2051852494607152128 &   12.67 & 12.65 &	4850.6	$\pm$12.5&	2.91$\pm$0.03 &-0.045$\pm$0.001 & 1.79 & \multirow{2}{*}{18008.3$\pm$676.1} \\
2449570 &   AB1-B & 2051852490306075648 &   12.71 & 12.70 &  	4769.9	$\pm$8.3&	2.92	$\pm$0.02  & 0.054$\pm$0.007   & 0.95
 \\
 \midrule
2570370 &   AB2-A & 2051295351447963136 & 11.67 & 11.63 &  4732.5	$\pm$14.1	&2.50	$\pm$0.03  & 0.192$\pm$0.012      &  0.99    & \multirow{2}{*}{19386.3$\pm$8.0}    \\
2570384 &   AB2-B & 2051295351447969152 & 14.99 & 14.96 &   	4519.4	$\pm$10.5	&2.69	$\pm$0.03  &0.302$\pm$0.011    &  1.01        \\
 \midrule
3122185 &   AB3-A & 2051928738868317440 &  11.18 & 11.16 &    4979.6	$\pm$11.6	&2.94	$\pm$0.02  & 0.146$\pm$0.008    &  1.00   & ...    \\
... & AB3-B & 2051928738864291072 &  ... & 12.20 & 4973.6	$\pm$9.6	&2.72	$\pm$0.02 &-0.127$\pm$0.007 & 1.68 & 12602.5$\pm$167.1\\
3122188 & ... & 2051928734562919168 &  11.11 & 11.62 &   $\mathrm{{5620^{+75}_{-84}}^{*}}$&    $\mathrm{3.43\pm0.27^{*}}$ & $\mathrm{0.06\pm0.15^{*}}$ & 	
3.55 & 13182.5$\pm$2.6 \\[2pt]
 \midrule
4042300 &   AB4-A & 2100394867817258624 &  10.73 & 10.74 &  	4752.5	$\pm$13.3	&2.52	$\pm$0.03 & -0.027$\pm$0.012     &  1.06
    & \multirow{2}{*}{6685.0$\pm$1.3}  \\
4042308 &   AB4-B & 2100394863523663872 &  12.13 & 12.14 & 	4695.5	$\pm$7.9	&2.46	$\pm$0.02 &  -0.166$\pm$0.008       & 1.27  \\
 \midrule
5098143 &   AB5-A & 2053444995457236736 &  12.96 & 12.92 &  4552.6	$\pm$11.8	&2.28	$\pm$0.03 & 0.142$\pm$0.012        &  0.97  & \multirow{2}{*}{16443.5$\pm$79.2}  \\
5098145 &   AB5-B & 2053444999760661120 &  13.11 & 13.08 &  	4667.0$\pm$12.7	&2.54	$\pm$0.03 & 0.143$\pm$0.012       & 0.95        \\
   \midrule
6430834 &  AB6-A & 2102442673866383744 &  10.67 & 10.63 & 4737.1$\pm$7.7	&2.49	$\pm$0.02  &  0.057$\pm$0.007	    &  0.87    & \multirow{2}{*}{6056.0$\pm$1.2}    \\
6430841 &  AB6-B & 2102442673865162112 &  11.16 & 11.13 &  4885.6 $\pm$ 9.4	&2.40	$\pm$0.02  &-0.307$\pm$0.007       &  0.84     \\
\hline \hline
\end{tabular}}
\tablefoot{
\textit{Gaia} DR3 ID, and \textit{G}mag (\textit{Gaia G}mag) are from \citet{2023_Martinez_Palomera}. Kp indicates the \textit{Kepler} magnitude. $T_\mathrm{eff}$, log $g$ and [Fe/H] are adopted from APOGEE DR17 values for the corresponding \textit{Gaia} DR3 IDs, except where indicated with *, which are from \citet{2015_Luo}. Projected physical separations ($s$) are measured with respect to the brightest star in \textit{G}mag of each AB candidate. For AB3, we indicated \textit{s} for both the separation between AB3-A and AB3-B, and between AB3-A and KIC 3122188.}
\label{tab:obs_info}
\end{table*}

\section{Measuring global properties of the oscillations} \label{sec:method}
\subsection{The background model and $\nu_{max}$ estimation} \label{subsec:background}
In the power density spectra of red-giant stars, stellar activity, granulation, photon and instrumental noise form a background on which the oscillations are superimposed. Instrumental noise is negligible \citep{2014_Kallinger}, thus the noise can be assumed to be white noise. It is necessary to remove these background signals to study oscillations. We used TACO (Tools for the Automated Characterisation of Oscillations, Hekker et al., in prep.) code to fit the background and obtain $\rm{\nu_{max}}$. We follow the description by \citet{2014_Kallinger}, to model the global background of the PDS ($\rm{P_{bg}(\nu)}$) as 
\begin{equation}
\rm{P}_{bg}(\nu) = \it{w}_{{\textrm{noise}}} + \rm{\eta}^2(\nu)\left(\sum_{i=1}^{3}\frac{A_i}{1+(\nu/b_i)^4}\right),   \label{eq:background}
\end{equation}
where $w_{\textrm{noise}}$ is the constant white noise level. The second term represents granulation at different scales as the sum of three super-Lorentzian components. For each component, $\rm{A_i}$ is the characteristic amplitude, and $\rm{b_i}$ is the characteristic (turnover) frequency (see e.g. Fig. \ref{fig:2570370_2570384_PDS}(b)). The apodization factor $\rm{\eta(\nu)}$ accounts for the finite sampling of the flux measurements in the light curves (for details see \citealt{2014_Kallinger}).

We fit a Gaussian function to account for the oscillation power excess in the PDS and estimate the frequency of maximum oscillation power, $\rm{\nu_{max}}$. The power excess was modeled as
\begin{equation}
\rm{P}_{{osc}}(\nu) = P_{\rm{g}}\exp \left(- \frac{(\nu-\nu_{\rm max})^2}{2\sigma_\mathrm{env}^2}\right)
\label{eq:oscillation}
,\end{equation}
where $\rm{P}_{\rm g}$, and $\rm{\sigma_{env}}$ are the height, and the width of the Gaussian envelope, respectively. We applied a Bayesian Markov chain Monte Carlo framework with affine-invariant ensemble sampling (emcee; \citealt{2013_Foreman_emcee}) to estimate all parameters of a global fit to the PDS: 
\begin{equation}
\rm{P(\nu) = P_{bg}(\nu) + \eta^2(\nu)P_{osc}(\nu)},
\label{eq:global-fit}
\end{equation}
where we define $\mathrm{P(\nu)}$ as the full PDS model. Subsequently, we used the medians of the posterior probability distributions for each parameter as estimates of the expectation values, with uncertainties given by the 16th and 84th percentiles. We calculated the background-normalized PDS to focus on the oscillation power by dividing the observed spectrum $\rm{P_{obs}(\nu)}$ by the fitted background model $\rm{P_{bg}(\nu)}$ (see e.g. Fig. \ref{fig:2570370_2570384_PDS}(c)).

\subsection{Mode identification and $\Delta\nu$ determination} \label{subsec:modeID}
We followed the method described by \citet{2018_Garcia} to automatically identify significant peaks in the background-normalized PDS. If a peak is an oscillation mode with a width larger than the frequency resolution, the individual oscillation mode in the power excess is well characterized by a Lorentzian function,
\begin{equation}
\rm{
P_{\rm peak}(\nu) = \frac{H_{\rm peak}}{1 + \left(\frac{\nu - \nu_{\rm peak}}{\gamma_{\rm peak}}\right)^2},
}
\label{eq:Lorentzian}
\end{equation}
where $\nu_{\rm peak}$, $\mathrm{H_{\rm peak}}$, and $\gamma_{\rm peak}$ are the mode frequency, mode height, and the half-width at half-maximum (half of the linewidth), respectively. The mode amplitude (A) of a Lorentzian profile is defined as the area beneath the mode profile, $\mathrm{A=\sqrt{\pi \mathrm{H\gamma}}}$. 

When the oscillation mode lifetime (damping time) is longer than the time-span of the timeseries data, the mode profiles are too narrow to be resolved within the width of the frequency bin. Therefore, the signal resembles an undamped sine wave, and is well described by the $\mathrm{sinc}^2$ function: 
\begin{equation}
\rm{
P_{\rm peak}(\nu) = H_{\rm peak}{ sinc^2}{\left(\pi\frac{\nu - \nu_{\rm peak}}{\delta\nu}\right)}, 
}
\label{eq:sinc}
\end{equation}
where $\delta\nu$ is the frequency resolution. In this case, the mode amplitude is $\mathrm{A = \sqrt{\pi \rm{H}\delta\nu}}$. 

We used the identified peaks as initial values to optimize parameters using maximum likelihood estimation (MLE). After fitting the modes modeled with Eq. \ref{eq:Lorentzian}, we examine the residual from the background-normalized PDS and flag any single frequency bin whose power is too high to be explained by the noise (false-alarm probability lower than $10^{-4}$). We fit these flagged modes with both Eqs. \ref{eq:Lorentzian} and \ref{eq:sinc} and the model with the lower Akaike Information Criterion (AIC) is chosen. 

For individual stars, we provide values $\nu_{\rm peak}$, amplitudes, $\gamma_{\rm peak}$, $\mathrm{H_{\rm peak}}$, and spherical degrees of the identified peaks, which are available online via the CDS website. Uncertainties are derived from the covariance matrix estimated by the inverse of the Hessian matrix. The reported AIC values correspond to the difference in AIC values between the model with and without the oscillation peak  ($\Delta \mathrm{AIC=AIC_{without}-AIC_{with}}$) while maintaining the other components of the model unchanged. When $\Delta \mathrm{AIC}> 0$, the model with the oscillation peak is preferred and higher $\Delta \mathrm{AIC}$ values implies stronger evidence. We selected the peaks with $\Delta\mathrm{AIC\ge2}$. 

Based on the universal pattern \citep{2011_Mosser} for solar-like oscillations, we assigned the spherical degrees of the modes. With the initial guess of $\mathrm{\Delta\nu}$ calculated from the scaling relation \citep{2010_Hekker}, we cross-correlated the observed PDS with the universal pattern within the frequency range $\nu_{\mathrm{max}}\pm\Delta\nu$ and locate the central $\ell=0$ mode closest to $\nu_{\mathrm{max}}$. Then iteratively moved further away from  $\nu_{\mathrm{max}}$ to identify $\ell=0$ modes in different radial orders. For each radial order, we also identified quadrupole ($\ell=2$) and octupole ($\ell=3$) modes within the expected frequency windows relative to the radial modes, using the small frequency separation ($\mathrm{\delta\nu_{02}}$ from \citealt{2010_Huber}, and $\mathrm{\delta\nu_{03}}$ from \citealt{2011_Mosser}). After identifying $\ell=0,2,\text{and }3$, any remaining significant peaks were classified as dipole ($\ell=1$) mode candidates. 

Finally, we determined $\Delta\nu$ as the slope of a weighted least-squares fit to the identified $\ell=0$ mode frequencies versus radial order, considering their frequency uncertainties. By dividing the PDS into segments of length $\mathrm{\Delta\nu}$ and stacking them, we created the frequency {\'e}chelle diagrams that allow us to verify the determined  $\mathrm{\Delta\nu}$ values (see examples in Fig. \ref{fig:2570370_2570384_frequency_echelle}). We also simultaneously determined the phase term of the pressure modes, $\mathrm{\epsilon_p}$, by this linear fit method. With the three central $\ell=0$ frequencies, we obtained central (local) $\mathrm{\Delta\nu}$ and $\mathrm{\epsilon_p}$ ($\mathrm{\Delta\nu_c}$ and $\mathrm{\epsilon_{p,central}}$, respectively), which we used to distinguish the evolutionary stage of the red-giant stars following the method from \citet[][see also Fig. \ref{fig:appendix_ES}(a)]{2012_Kallinger}. 

\subsection{Characterization of the mixed-mode parameters} \label{subsec:deltaP_q} 

We implemented the Bayesian Optimisation for the Characterisation of Mixed Modes (BOChaMM\footnote{{https://github.com/jsk389/BOChaMM}}) pipeline, which performs forward modelling of mixed mode frequencies to infer the mixed-mode parameters (see \citealt{2023_Kuszlewicz} for details). We used the background-normalized PDS and focused on the mode frequencies of $\ell=1$ modes identified by TACO in the previous section as input mixed-mode frequencies. We analyzed stars for which TACO identified at least three $\ell=1$ modes per radial order. 

The forward modelling in BOChaMM is based on the asymptotic expansion presented for the mixed mode frequencies ($\nu$) by \citet{2012_Mosser_core}: 
\begin{equation} \label{eq:asymptotic_mixed_modes}
    \nu = \nu_{n_{p,l}} + \frac{\Delta\nu}{\pi} \arctan \bigg[q_l\tan\pi\bigg(\frac{1}{\Delta\Pi_l\nu}-\epsilon_g\bigg)\bigg],
\end{equation}
where $\nu_{n_{p,l}}$ is the underlying $p$-mode frequency of radial order $n_p$ and degree $l$, to which the $g$-mode is coupled, $\Delta\nu$ is the large frequency separation between $p$-modes, $\Delta\Pi_l$ is the asymptotic period spacing of modes with degree $l$, $q_l$ is the coupling factor between the $p$ and $g$ modes (with $0\leq q_l \leq1$, and $q_l$ = 0 means no coupling), and $\epsilon_g$ is the $g$-mode phase offset. Following \citet{2015_Mosser}, the pipeline constructs stretched periods ($\tau$) by integrating  
\begin{equation}\label{eq:stretched_period}
\mathrm{d}\tau = \frac{1}{\zeta(\nu)}\frac{\mathrm{d}\nu}{\nu^2},
\end{equation}
so that the $\ell=1$ mixed modes become nearly evenly spaced in $\tau$ with spacing of $\Delta\Pi_1$. $\zeta(\nu)$ quantifies how much the observed mixed-mode period spacings deviate from the asymptotic value of $\Delta\Pi_1$, hence $g$-dominated mixed modes have $\zeta\simeq1$, while $p$-dominated mixed modes have smaller $\zeta$. The pipeline adopts the expression for $\mathrm{\zeta(\nu)}$ given by \citet{2017_Hekker}. 

The BOChaMM pipeline first optimizes the signal in the power spectrum of the stretched-period spectrum, which is defined as $\mathrm{{PS \times PS}}$ in \citet{2023_Kuszlewicz}. Then the pipeline returns the initial estimates of the asymptotic dipole mode period spacing ($\mathrm{\Delta\Pi_{1,{PS \times PS}}}$), and coupling factor (${q_\mathrm{PS \times PS}}$) from the $\mathrm{{PS \times PS}}$.\textcolor{black}{\footnote{\textcolor{black}{For this step, we adopted broad uniform priors on $\Delta\Pi_1$ following \citet{2023_Kuszlewicz}, $20~\mathrm{s} \le \Delta\Pi_1 \le 200~\mathrm{s}$ for RGB and $50~\mathrm{s} \le \Delta\Pi_1 \le 400~\mathrm{s}$ for CHeB stars. In both cases we set the initial $q$ to $q_{\rm init}=0.2$, same as the original pipeline and performed $2000$ iterations.}}} Subsequently, $\mathrm{\Delta\Pi_{1,{PS \times PS}}}$ is passed forward as the input for the Bayesian search over $\mathrm{\Delta\Pi_{1}}$, $\mathrm{\text{q}}$,  $\mathrm{\epsilon_g}$, and $\mathrm{\delta\nu_{rot}}$ by fitting the observed mixed mode frequencies. The $\mathrm{\delta\nu_{rot}}$ indicates the frequency splittings of mixed modes due to the effect of rotation, assuming that three rotational components are present in the mixed-modes ($m = 0, \pm1$, where $m$ is the azimuthal order). For each star, we iterated 7500 times to obtain the best-fit values. 

\subsection{Fundamental stellar parameters from asteroseismic scaling relations} \label{subsec:scaling} 
\citet{1991_Brown} suggested a scaling relation between $\nu_{\mathrm{max}}$ and the acoustic cut-off frequency ($\mathrm{\nu_{ac}}\propto g{T_\mathrm{eff}}^{-\frac{1}{2}}$), with $g$ and $T_\mathrm{eff}$ being the surface gravity and the effective temperature, respectively. Therefore, $\nu_{\mathrm{max}}$ can be expressed as (\citealt{1995_Kjeldsen}): 
\begin{equation}
\begin{aligned}
\nu_{\max}&\simeq \frac{M/M_\odot}{\left(R/R_\odot\right)^2 \sqrt{T_{\mathrm{eff}}/T_{\mathrm{eff,\odot}}}} 
\nu_{\max,\mathrm{\odot}},
\label{eq:numax}
\end{aligned}
\end{equation}
with $M$ mass, and $R$ radius. The large frequency separation is related to the inverse sound travel time across the stellar diameter and is therefore a measure of the mean stellar density $\mathrm{\bar\rho}$ (\citealt{1986_Ulrich}): 
\begin{equation}
\begin{aligned}
\Delta\nu &\simeq
\left(\frac{M}{M_\odot}\right)^{1/2}
\left(\frac{R}{R_\odot}\right)^{-3/2}
\, \Delta\nu_{\mathrm{\odot}} \\
&\simeq
\sqrt{\bar{\rho}/\bar{\rho}_\odot}
\Delta\nu_{\mathrm{\odot}},
\end{aligned}
\label{eq:deltanu}
\end{equation}
By combining these two scaling relations (Eqs. \ref{eq:numax} and \ref{eq:deltanu}), we can compute stellar parameters from the observed oscillations with the Sun as the reference star: 
\begin{equation}
\begin{aligned}
\frac{M}{M_\odot}
&\simeq
\left(\frac{\nu_{\max}}{f_{\nu_{\max}}\,\nu_{\max,\odot}}\right)^{3}
\left(\frac{\Delta\nu}{f_{\Delta\nu}\,\Delta\nu_\odot}\right)^{-4}
\left(\frac{T_{\mathrm{eff}}}{T_{\mathrm{eff},\odot}}\right)^{3/2},
\end{aligned}\label{eq:mass_corr}
\end{equation}
\begin{equation}
\begin{aligned}
\frac{R}{R_\odot}
&\simeq
\left(\frac{\nu_{\max}}{f_{\nu_{\max}}\,\nu_{\max,\odot}}\right)
\left(\frac{\Delta\nu}{f_{\Delta\nu}\,\Delta\nu_\odot}\right)^{-2}
\left(\frac{T_{\mathrm{eff}}}{T_{\mathrm{eff},\odot}}\right)^{1/2},
\end{aligned}\label{eq:radius_corr}
\end{equation}
where $f_{\nu_{\max}}$ and ${f_{\Delta\nu}}$ are correction factors for the $\nu_{\mathrm{max}}$ and $\Delta\nu$ scaling relations, respectively. We adopted solar reference values from \citet{2018_Themessl_EB}, with $\mathrm{\nu_{max,\odot}=3166\pm6 \mu Hz}$, $\mathrm{\Delta\nu_{\odot} =135.4\pm0.3 \mu Hz}$, and ${T_\mathrm{eff,\odot}=5772.0\pm0.8\mathrm{K}}$ (\citealt{2015_Mamajek,2016_Prsa}). For each star, we used $T_\mathrm{eff}$ from APOGEE DR17 \citep{2022_Abdurrouf}. 

By using the solar value as a reference in the $\Delta\nu$ scaling relation (Eq. \ref{eq:deltanu}), we inherently assume that the structure of a red-giant star is homologous to the Sun. This is not entirely the case due to the high contrast in the density between the core and the envelope in red-giant stars (\citealt{2011_White}, and see review by \citealt{2020_Hekker}). Therefore, we applied $ {f_{\Delta\nu}}$ which we computed from Asfgrid (\citealt{2016_Sharma, 2022_Stello_Sharma}). The factor $ {f_{\Delta\nu}}$ is a function of the metallicity, $T_\mathrm{eff}$, log $g$, mass and the evolutionary phase. For the metallicity and log $g$, we used APOGEE DR17 \citep{2022_Abdurrouf} values. 

In principle, the factor $f_{\nu_{\max}}$ can be calibrated empirically by comparing the seismically inferred radii with \textit{Gaia} radii which depend on parallax (e.g. \citealt{2025_Pinsonneault}). If oscillation signals from multiple stars overlap, the measured $\mathrm{\nu_{\max}}$ can be biased, which in turn biases the asteroseismic radius inferred from the scaling relations. This leads to unreliable values of $f_{\nu_{\max}}$. Since our goal is to compare the stellar parameters derived from the same equations (eqs. \ref{eq:mass_corr} and \ref{eq:radius_corr}) between the PDS with seismically unresolved oscillations and the PDS of the individual stars, we fix $f_{\nu_{\max}}=1$ in this work. Finally, the uncertainties in mass and radius were calculated by propagating the uncertainties in $\nu_{\mathrm{max}}$, $\Delta\nu$, and $T_\mathrm{eff}$.

\section{Results} \label{sec:Results}
Seismically unresolved red-giant AB candidates selected in Sect. \ref{subsec:identify_AB} show diverse PDS morphologies. The morphology of these PDSs varies depending on the flux contributions and the PDS morphologies of the two stars. Some resemble that of a single star, while others appear distinct and show complex oscillation patterns. These characteristics provide useful diagnostics for identifying them in the observations. In Sect. \ref{subsec:PDS}, we compare the PDS morphology of the identified AB candidates with that of the individual stars. We analyze asteroseismic and fundamental stellar parameters in Sect. \ref{subsec:result_seis_params}, and subsequently compare mixed mode parameters derived from the PDSs of AB candidates with those from the individual stars in Sect. \ref{subsec:result_dipole}. 

\subsection{PDSs of seismically unresolved AB candidates vs. individual stars} \label{subsec:PDS}

\begin{figure}[!htbp]
    \centering
    \makebox[\linewidth][l]{
        \includegraphics[width=0.95\linewidth]{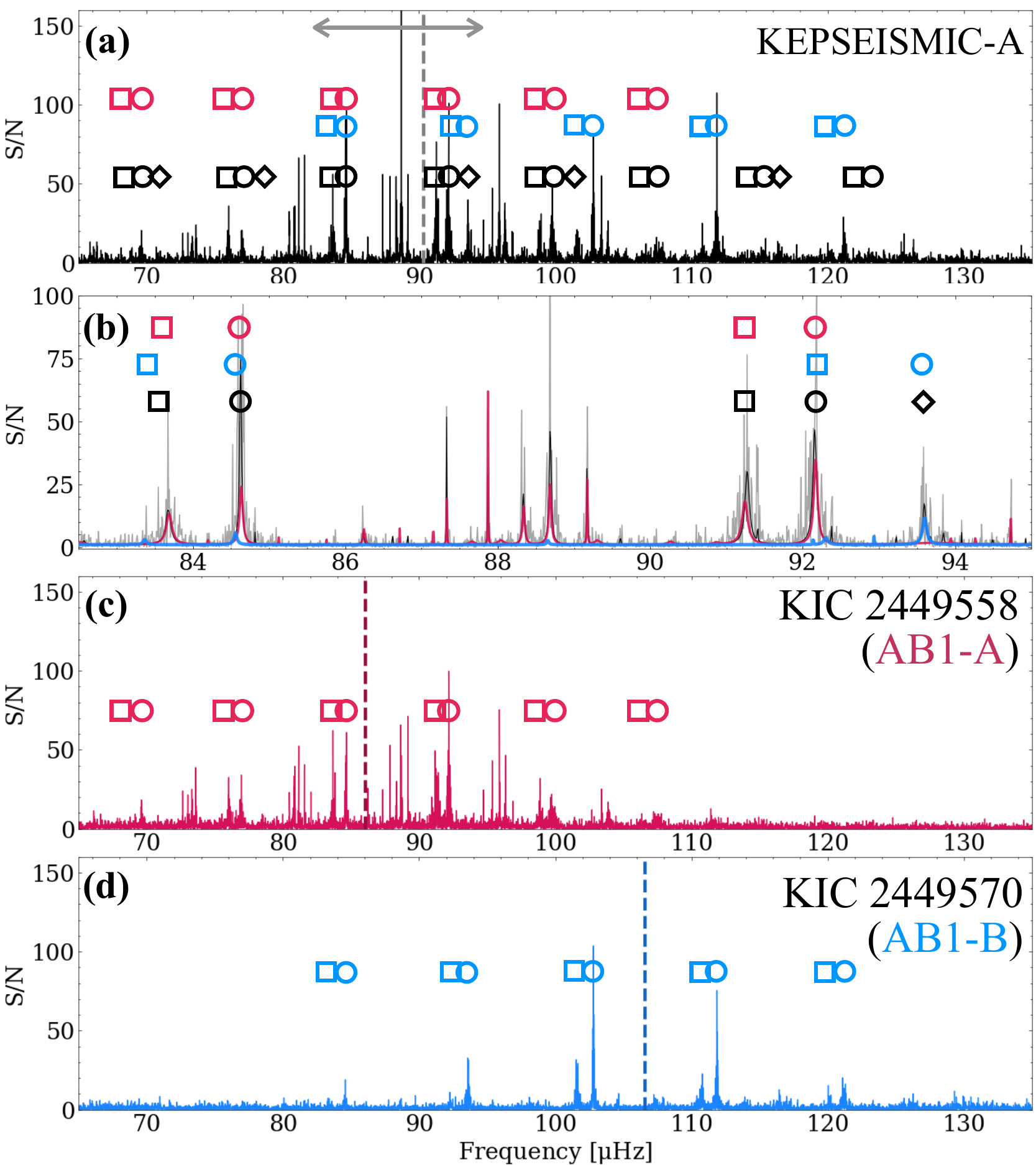}}
    \caption{Background-normalized PDSs of AB1, AB1-A and AB1-B. Panel (a) shows overlapped power excesses from both stars (from the KEPSEISMIC light curve of AB1-A). The gray arrow marks the region shown in panel (b). In panel (b), the peaks in gray represent the data, and the red and blue lines are fits for AB1-A and AB1-B. The PDSs in panels (c) and (d) are derived from KBonus light curves. The vertical dashed lines in panels (a), (c), and (d) indicate $\mathrm{\nu_{max}}$. The modes are marked by circles ($\mathrm{\ell=0}$), squares ($\mathrm{\ell=2}$), and diamonds ($\mathrm{\ell=3}$). The dipole modes are not marked.}
    \label{fig:PDS_2449}
\end{figure}

\textbf{\textit{KIC 2449558 (AB1-A) and KIC 2449570 (AB1-B):}} AB1-A and AB1-B have similar brightness ($\mathrm{Kp}$ =12.67 and 12.71, respectively). Therefore, we expect to detect oscillations from both stars in the PDS when the aperture captures a similar fraction of flux from each star. As shown in Fig. \ref{fig:PDS_2449}(a), the oscillation peaks from both stars are clearly visible and they overlap between $\mathrm{\nu\sim 80}$ and 110 $\mathrm{\mu Hz}$. Their superposition broadens the combined power excess relative to that of the individual stars shown in Figs. \ref{fig:PDS_2449}(c) and (d). However, the overall PDS morphology remains similar to that of AB1-A. This is because the $\mathrm{\ell=0}$ and $\mathrm{\ell=2}$ (i.e., $\mathrm{\ell=0,2}$ pair) modes of the two stars near $\mathrm{\nu\approx 85\mu Hz}$ (see Fig. \ref{fig:PDS_2449}(b)) have almost identical frequencies (i.e., they are aligned; see \citealt{2025_Choi} and Appendix \ref{sec:appendix-choi}). In addition, the dipole mode amplitudes of AB1-B are notably suppressed and this star has been identified as a suppressed dipole mode star by \citet{2016_Stello_Nature} (see also \citealt{2012_Mosser, 2017_Mosser, 2024_Coppee}).

Nonetheless, an interesting configuration occurs near $\mathrm{\nu\approx 92\mu Hz}$ in Fig. \ref{fig:PDS_2449}(b). The $\mathrm{\ell=0}$ mode frequency from AB1-A aligns with the $\mathrm{\ell=2}$ mode frequency of AB1-B (i.e., partially aligned; see \citealt{2025_Choi}). As a result, the prominent peak at $\mathrm{\nu\approx 94\mu Hz}$ can be misidentified as an $\mathrm{\ell=3}$ mode in the PDS of AB1, even though it actually originates from the $\mathrm{\ell=0}$ mode of AB1-B. A similar feature appears near $\mathrm{\nu\approx 102\mu Hz}$ in Fig. \ref{fig:PDS_2449}(a). Finally, all identified peaks above 110$\mathrm{\mu Hz}$ in Fig.  \ref{fig:PDS_2449}(a) belong to AB1-B.

\textbf{\textit{KIC 2570370 (AB2-A) and KIC 2570384 (AB2-B):}} Figs. \ref{fig:2570370_2570384_PDS}(b), (e), (h) and (k) shows a single power excess in each panel, regardless of whether the underlying signals originate from a single star or from two stars. This highlights the difficulties of identifying seismically unresolved AB candidates. Moreover, the background-normalized PDS in Fig. \ref{fig:2570370_2570384_PDS}(c) does not allow us to confirm the presence of oscillation peaks from both stars due to the short time-span of the light curve shown in Fig. \ref{fig:2570370_2570384_PDS}(a). In contrast, the PDS shown in Fig. \ref{fig:2570370_2570384_PDS}(i) clearly exhibits oscillation signals from both stars. It shows a dense forest of peaks with complex oscillation patterns around $\mathrm{\nu\sim 50-60\mu Hz}$. Although TACO identified a few $\mathrm{\ell=0,2}$ pairs in Fig. \ref{fig:2570370_2570384_PDS}(i), these are only available below $\mathrm{\nu\sim 55\mu Hz}$ as shown in black symbols. Moreover, the frequency values of the identified  $\mathrm{\ell=0,2}$ pairs are inconsistent with the  $\mathrm{\ell=0,2}$ pair frequencies of either AB2-A or AB2-B.

\begin{figure}[!htbp]
    \centering
    \includegraphics[width=0.95\linewidth]{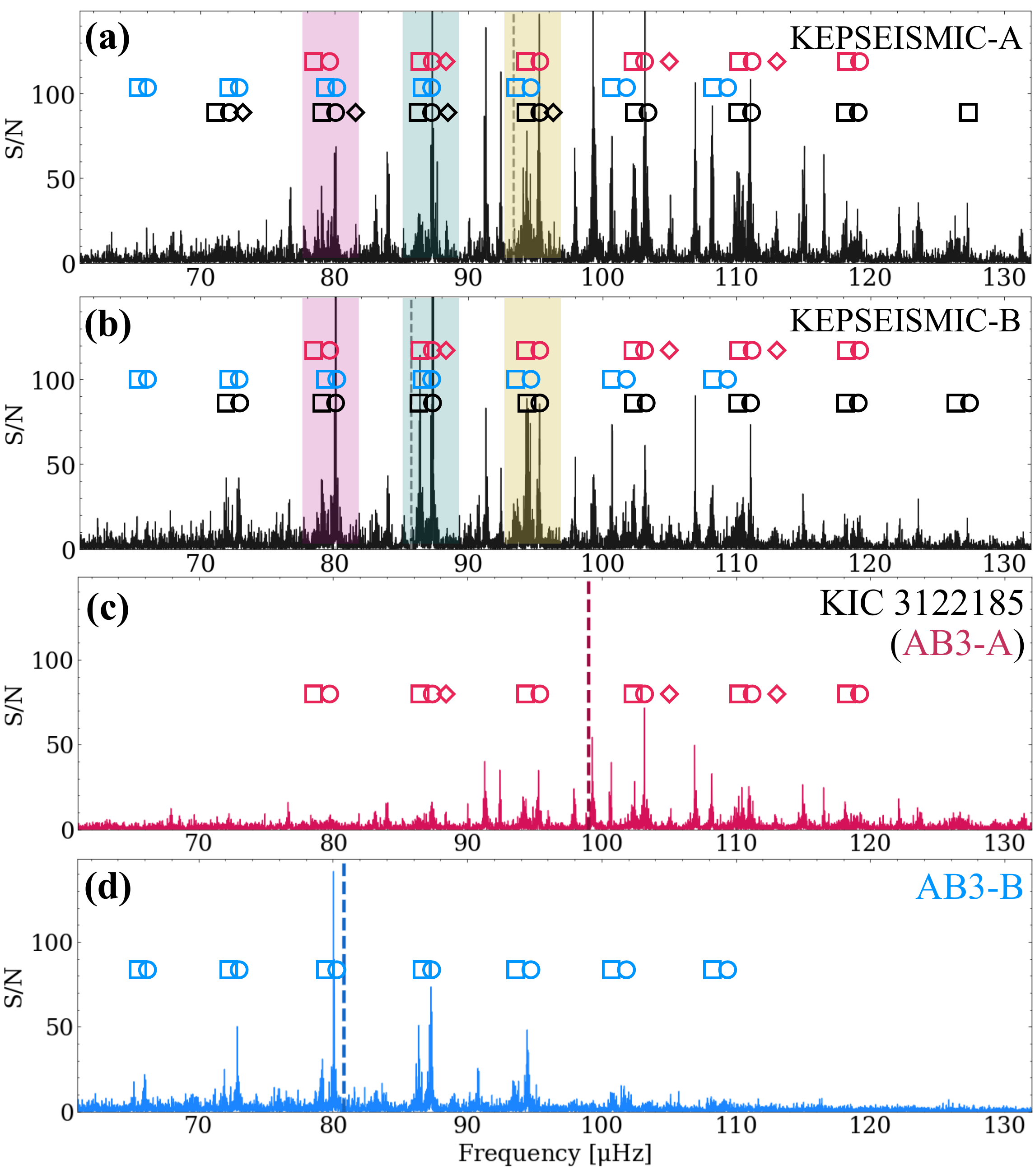}
    \caption{Similar as Fig. \ref{fig:PDS_2449}, now for the PDSs of AB3 (panels (a) and (b)), AB3-A and AB3-B. Panel (c) was derived from PDCSAP light curve, and panel (d) was derived from the custom light curve. The shaded regions in panels (a) and (b) indicate the frequency ranges where the $\mathrm{\ell=0,2}$ pairs from AB3-A and AB3-B mainly overlap. See main text for more details.}\label{fig:PDS_suppressed_3122}
\end{figure}

\textbf{\textit{KIC 3122185 (AB3-A) and \textit{Gaia} DR3 2051928738864291072 (AB3-B):}} Fig. \ref{fig:PDS_suppressed_3122} shows the PDSs of AB3 from two KEPSEISMIC light curves, which illustrates different relative flux contributions from the two stars. Panel (a) is derived from a light curve where the oscillation signal of AB3-A is more prominent, while panel (b) is derived from a light curve where AB3-B dominates the signal. Despite this difference, the two PDSs show oscillation peaks at similar frequencies because the blended $\mathrm{\ell=0,2}$ pairs do not create complex oscillation patterns. For example in Fig. \ref{fig:PDS_suppressed_3122}(a) and (b), the $\mathrm{\ell=0,2}$ pair of AB3-B dominates in the red-shaded region, and the $\mathrm{\ell=0,2}$ pair of AB3-A dominates in the yellow-shaded region. Moreover, the two $\mathrm{\ell=0,2}$ pairs align in the blue-shaded region. 

Due to the combined oscillations of AB3-A and AB3-B, Figs. \ref{fig:PDS_suppressed_3122}(a) and (b) show remarkably different PDS morphologies from the PDSs in panels (c) and (d). AB3-B in particular shows low amplitude dipole modes. If the analysis of the AB3-B is solely based on the PDS in Fig. \ref{fig:PDS_suppressed_3122}(a) or (b), peaks from AB3-A can be misinterpreted as dipole modes of the AB3-B. This leads to the overestimation of the $\mathrm{\ell=1}$ mode visibility and makes it difficult to recognize AB3-B as a star with suppressed dipole modes. 

\begin{figure}[!htbp]
    \centering
    \includegraphics[width=0.95\linewidth]{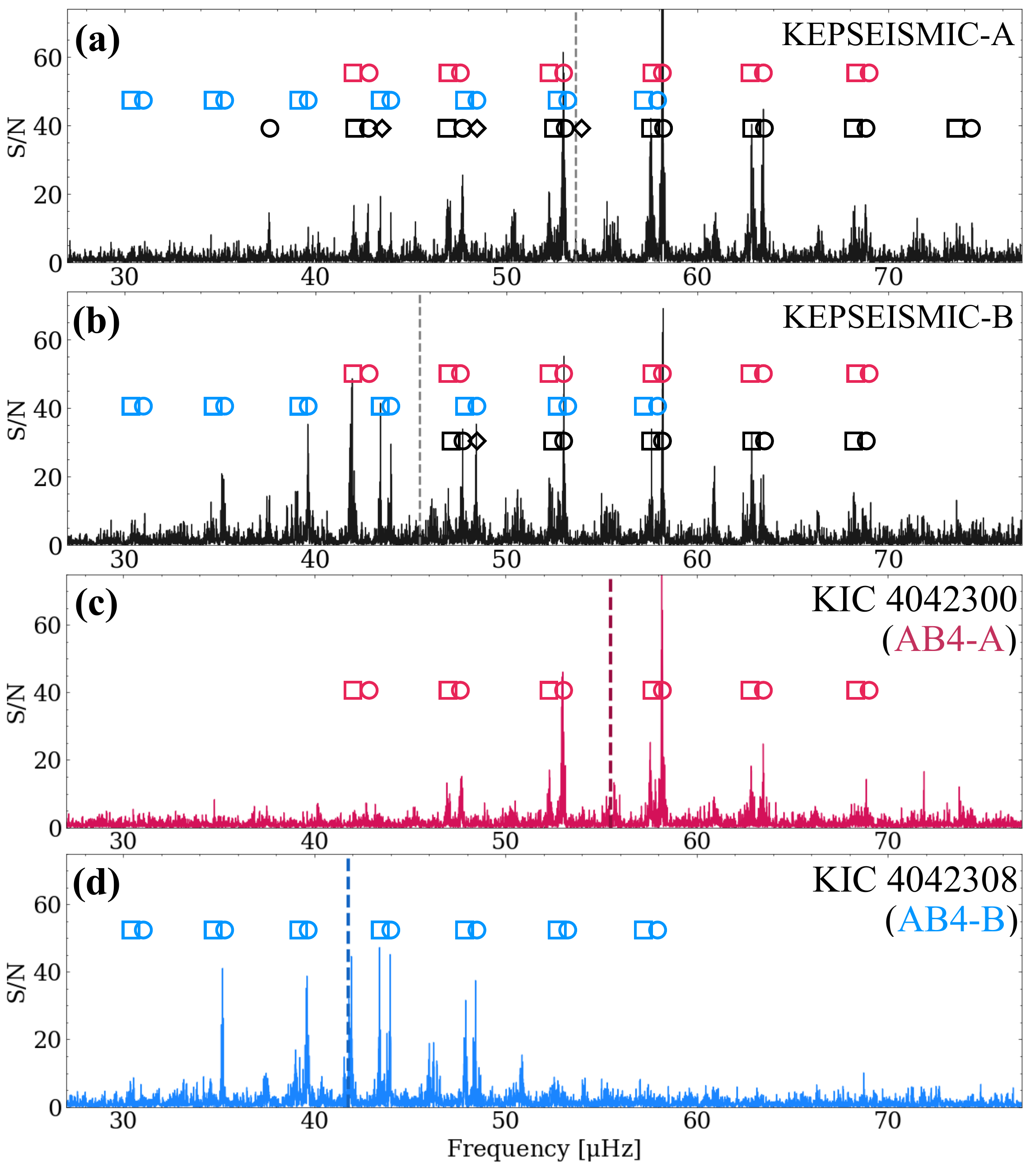}
    \caption{Similar as Fig. \ref{fig:PDS_2449}, now for the PDSs of AB4, AB4-A and AB4-B. Panels (c) and (d) were derived from PDCSAP light curves.}  \label{fig:PDS_suppressed_404}
\end{figure}

\textbf{\textit{KIC 4042300 (AB4-A) and KIC 4042308 (AB4-B):}} AB4 also includes a star with low-amplitude dipole modes (AB4-A; see Fig. \ref{fig:PDS_suppressed_404}(c)). Since the $\mathrm{\nu_{max}}$ values between the two stars differ by $\sim 30$\%, it is relatively easy to identify as an AB candidate from the PDS when both stars contribute comparable flux to the light curves. As shown in Fig. \ref{fig:PDS_suppressed_404}(b), the PDS shows a clear change in the oscillation pattern near $\mathrm{\nu\approx 50\mu Hz}$. Across this transition, the dominant peaks no longer follow the universal pattern with a defined sequence and this makes the identification of modes particularly challenging. TACO consequently identified the $\mathrm{\ell=0,2}$ pairs only at the higher frequency end of the spectrum ($\mathrm{\nu_{max}\gtrsim 45\mu Hz}$; see panel (b)). By contrast, the PDS in panel (a) closely resembles the PDS of AB4-A in panel (c). In that case, the identified modes agree well with AB4-A. 

\begin{figure*}[!htbp]
    \centering
    \includegraphics[width=1\linewidth]{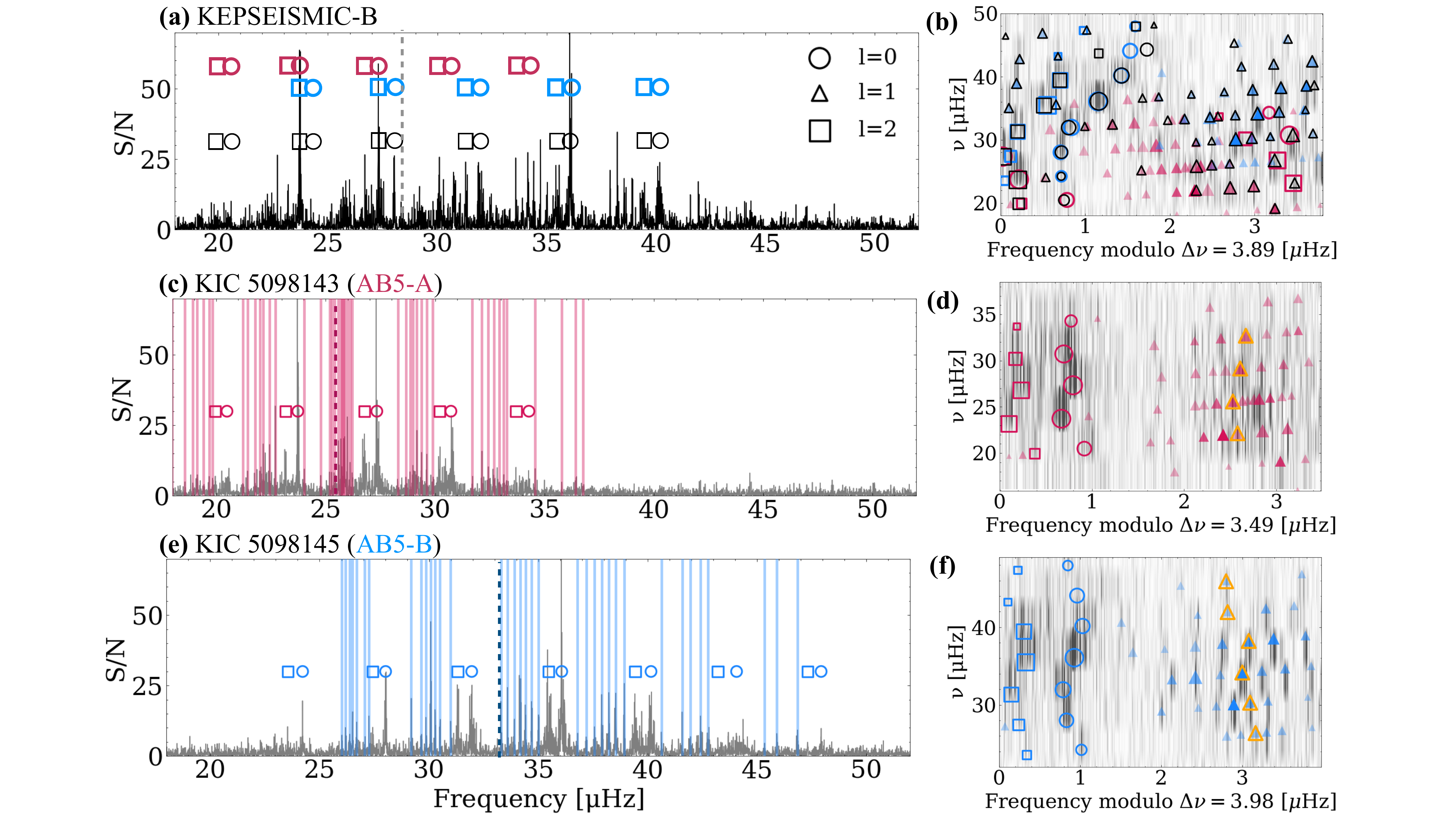}
    \caption{Background-normalized PDSs and frequency {\'e}chelle diagrams of AB5, AB5-A and AB5-B. Panels (a), (c) and (e): Similar to Fig. \ref{fig:PDS_2449}, mode degrees are labeled as per the legend. In panels (c) and (e), we used KBonus light curves and the vertical red and blue lines indicate $\mathrm{\ell=1}$ modes. Panels (b), (d) and (f): Frequency {\'e}chelle diagrams for the spectra on the left, computed with the $\mathrm{\Delta\nu}$ values listed below each panel. We highlight the modes with the lowest $\mathrm{\zeta}$ values by orange outline in panels (d) and (f).} 
    \label{fig:5098_pds}
\end{figure*}

\begin{figure*}[!htbp]
    \centering
    \includegraphics[width=1\linewidth]{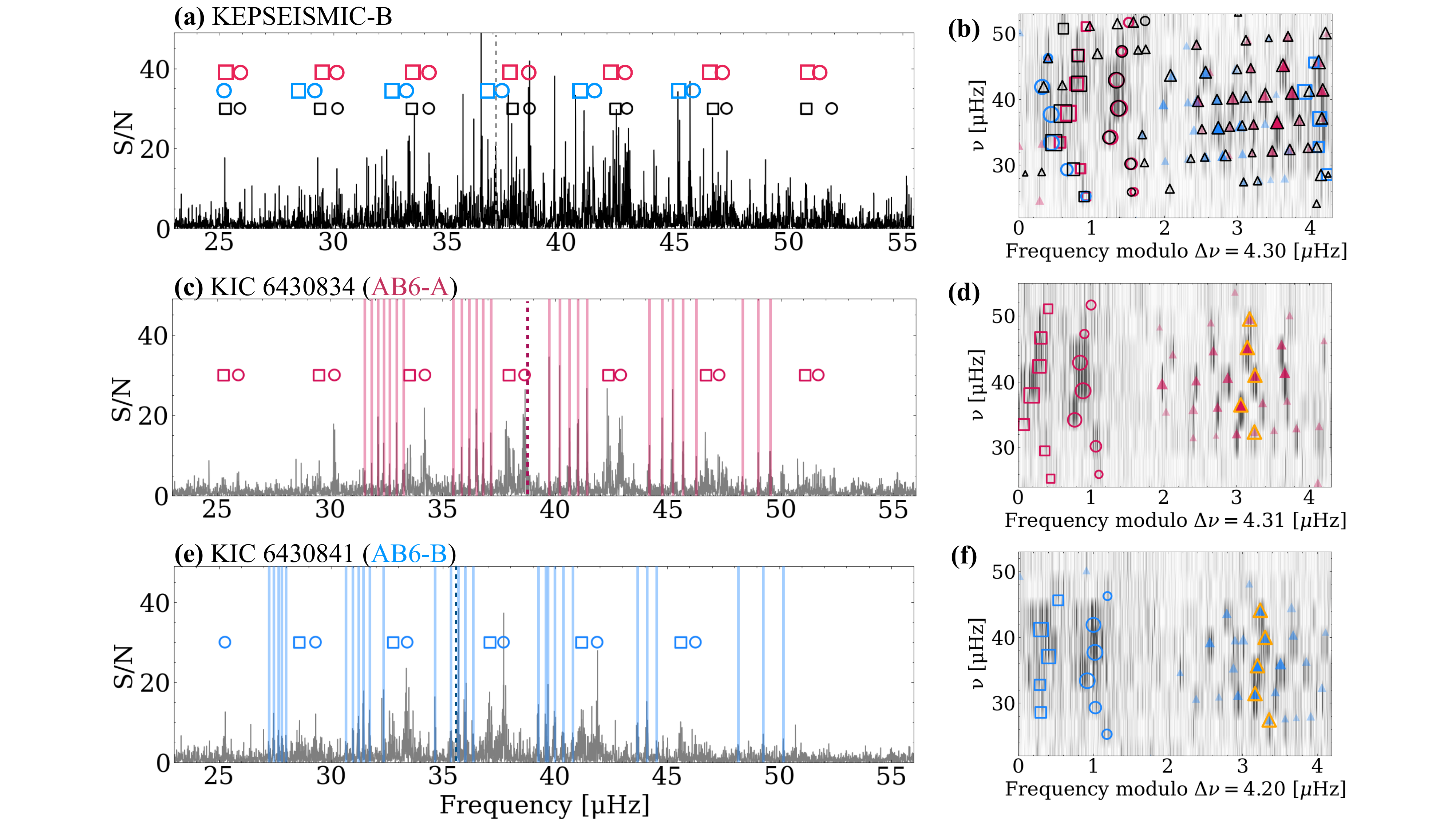}
    \caption{Similar to Fig. \ref{fig:5098_pds}, now for AB6, AB6-A and AB6-B. Panel (a) shows oscillation signals from both AB6-A and AB6-B in the PDS. We labeled mode degrees same as Figs. \ref{fig:5098_pds}. Panels (c) and (e) show PDSs derived from custom light curves.}
    \label{fig:6430834_PDS}
\end{figure*}

\textbf{\textit{KIC 5098143 (AB5-A) and KIC 5098145 (AB5-B):}} AB5 consists of two CHeB stars. The evolutionary states are determined from their $\mathrm{\Delta\nu}$ and central $\mathrm{\epsilon_p}$ values following the method from \citet{2012_Kallinger}. In Fig. \ref{fig:5098_pds}(a), the spectrum shows partially aligned frequencies at $\mathrm{\nu\approx 24\mu Hz}$ and $\mathrm{\nu\approx 27\mu Hz}$. In addition, the $\mathrm{\ell=0,2}$ pairs of the two stars appear at different frequencies above $\mathrm{\nu\sim 30\mu Hz}$ (i.e., misaligned; \citealt{2025_Choi}). Consequently, the PDS in Fig. \ref{fig:5098_pds}(a) shows a complex structure. Outside the overlapped region, both sides of the power excesses show distinct oscillation patterns from each star (AB5-A dominates on the lower frequency side, and AB5-B on the higher frequency side) due to their $\mathrm{\nu_{max}}$ difference ($\sim$ 27\%). The identified $\mathrm{\ell=0,2}$ pair marked with black symbols at $\mathrm{\nu\approx 20\mu Hz}$ in panels (a) and (b) of Fig. \ref{fig:5098_pds} matches with the $\mathrm{\ell=0,2}$ pair of AB5-A, while the remaining pairs follow the pattern of AB5-B. 

\textbf{\textit{KIC 6430834 (AB6-A) and KIC 6430841 (AB6-B):}} The PDS of AB6 shows the most complex oscillation pattern among the six AB candidates (see Fig. \ref{fig:6430834_PDS}(a)). Importantly, AB6 agrees well with the prediction from \citet{2025_Choi} for the most complex PDS configuration: two CHeB stars oscillating at similar frequency ranges with misaligned morphologies. The difference in the $\mathrm{\nu_{max}}$ values of AB6-A and AB6-B ($\sim8\%$) is the smallest among AB1-AB6. Most $\mathrm{\ell=0,2}$ pairs of AB6-A and AB6-B are misaligned in frequency, and the resulting PDS of AB candidate differs substantially from the oscillation pattern of each star. Under these circumstances, any mode identification performed on the blended spectrum is inherently incorrect and will further bias the inferred seismic properties. 

\subsection{Comparison of the asteroseismic and fundamental stellar parameters} \label{subsec:result_seis_params}

\begin{figure*}[!htbp]
    \centering
    \vspace*{-0.01\linewidth}
    \hspace*{-0.02\linewidth}
    \includegraphics[width=0.98\linewidth]{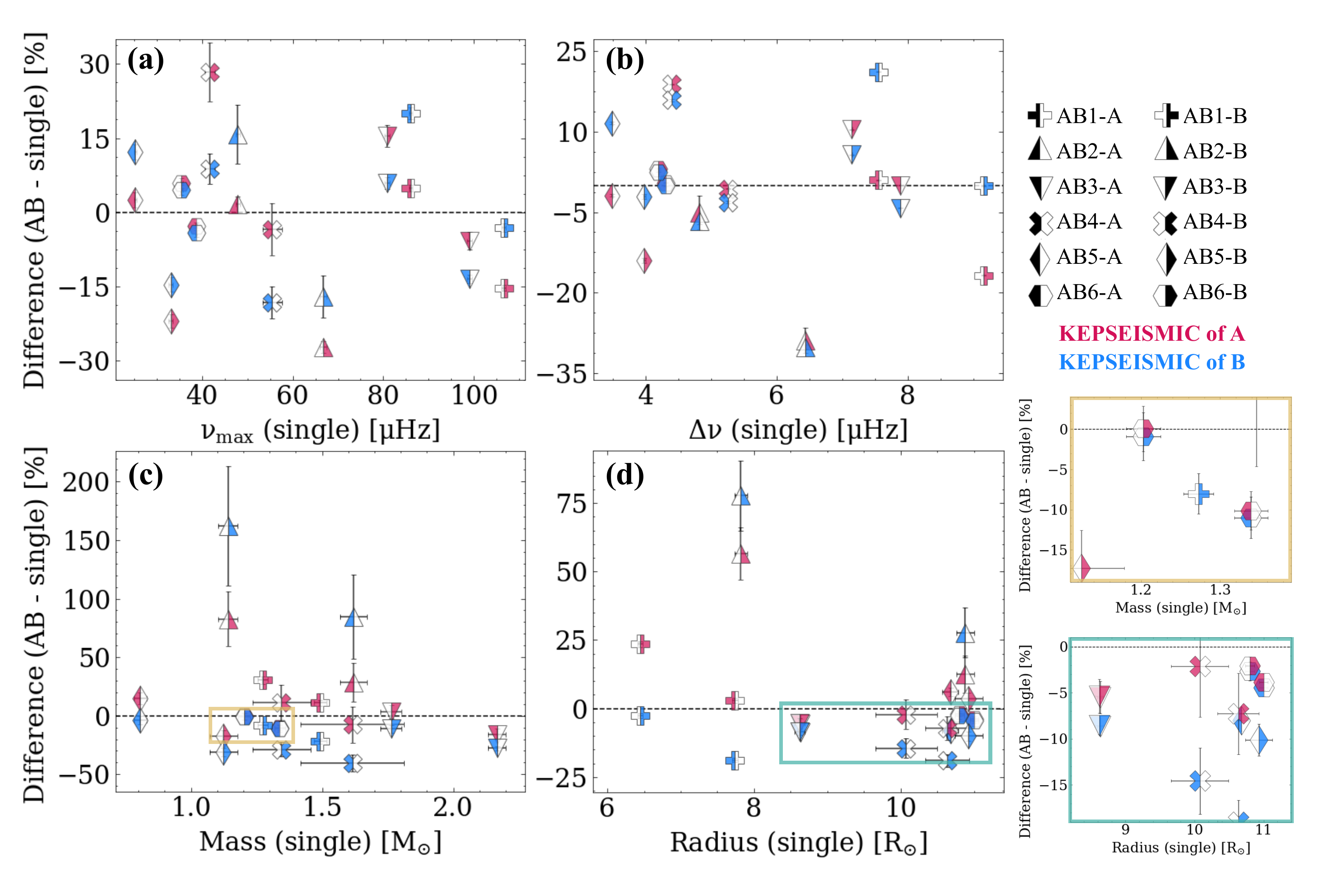}
    \caption{Percentage differences in $\mathrm{\nu_{max}}$, $\mathrm{\Delta\nu}$, mass, and radius derived from the PDSs of AB candidates and single stars ($\mathrm{(AB-single)\times100[\%]/single}$). The horizontal dashed lines at 0 indicate no difference between the values for AB candidate and single stars. Positive(negative) values imply that the parameters of the AB candidate are over(under)estimated relative to the single stars. Symbol shapes are defined in the legend and their colors indicate whether the KEPSEISMIC light curve of ABi-A or ABi-B (i=1-6) was used. The yellow and green boxes in the lower panels are zoomed in the right-hand panels to resolve the overlapping symbols.} 
    \label{fig:result_scatter}
\end{figure*}

\begin{figure}[!htbp]
    \centering
    \includegraphics[width=0.94\linewidth]{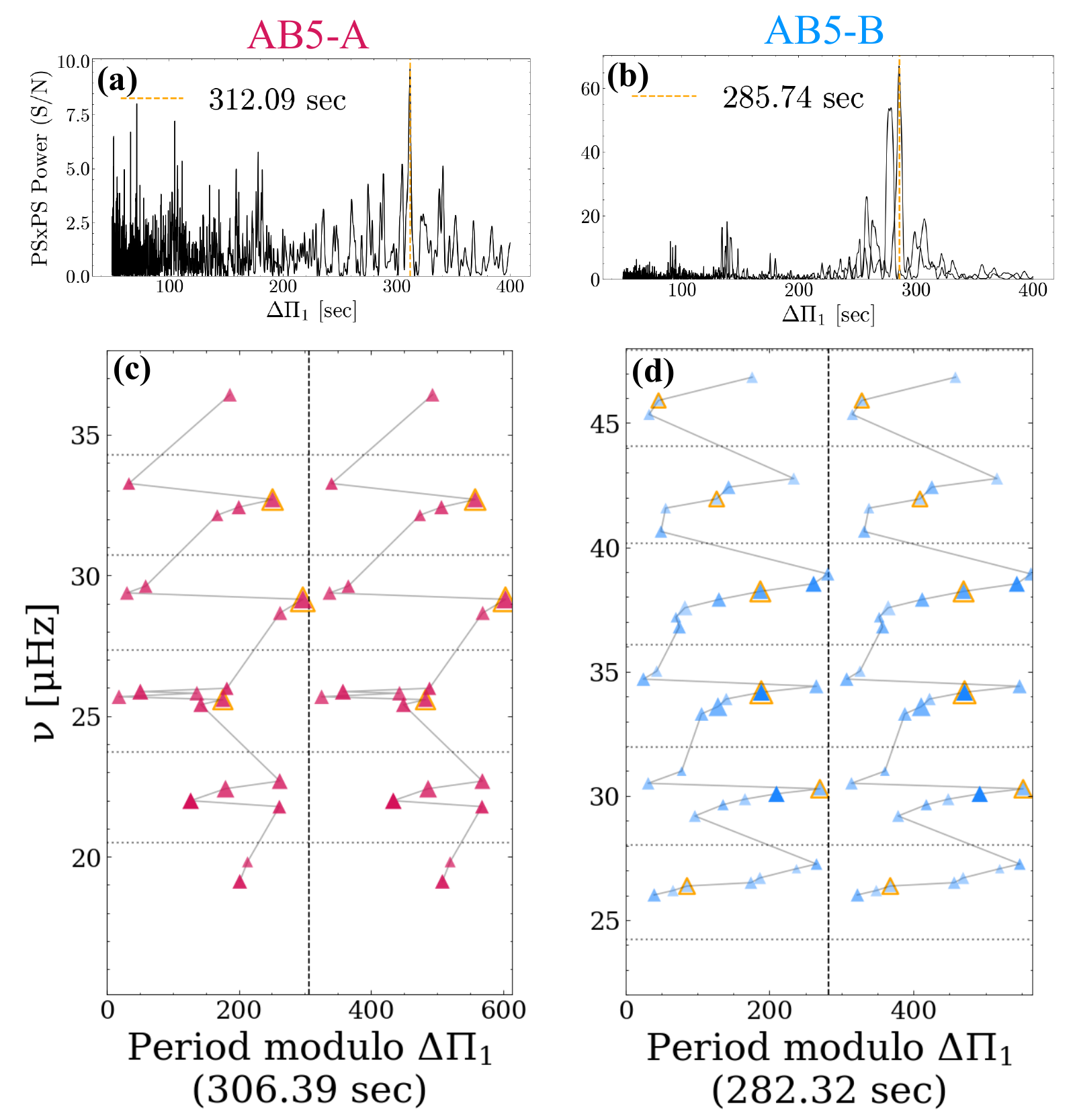}
    \caption{Dipole mixed mode analysis of KIC 5098143 (AB5-A) and KIC 5098145 (AB5-B). Panels (a) and (b): Fourier transform of the stretched period spectrum for $\mathrm{\ell=1}$ modes. The orange dashed lines indicate the highest power. Panels (c) and (d): Period {\'e}chelle diagram of $\mathrm{\ell=1}$ modes plotted twice and separated by the vertical dashed line which indicates the optimized $\mathrm{\Delta\Pi_1}$. Symbols for $\mathrm{\ell=1}$ modes are the same as Fig. \ref{fig:5098_pds}, and darker symbols correspond to higher $\Delta \mathrm{AIC}$ values. The horizontal dashed lines mark the radial mode positions.}
    \label{fig:5098_period}
\end{figure}

\begin{figure*}[!htbp]
    \centering
    \includegraphics[width=0.75\linewidth]{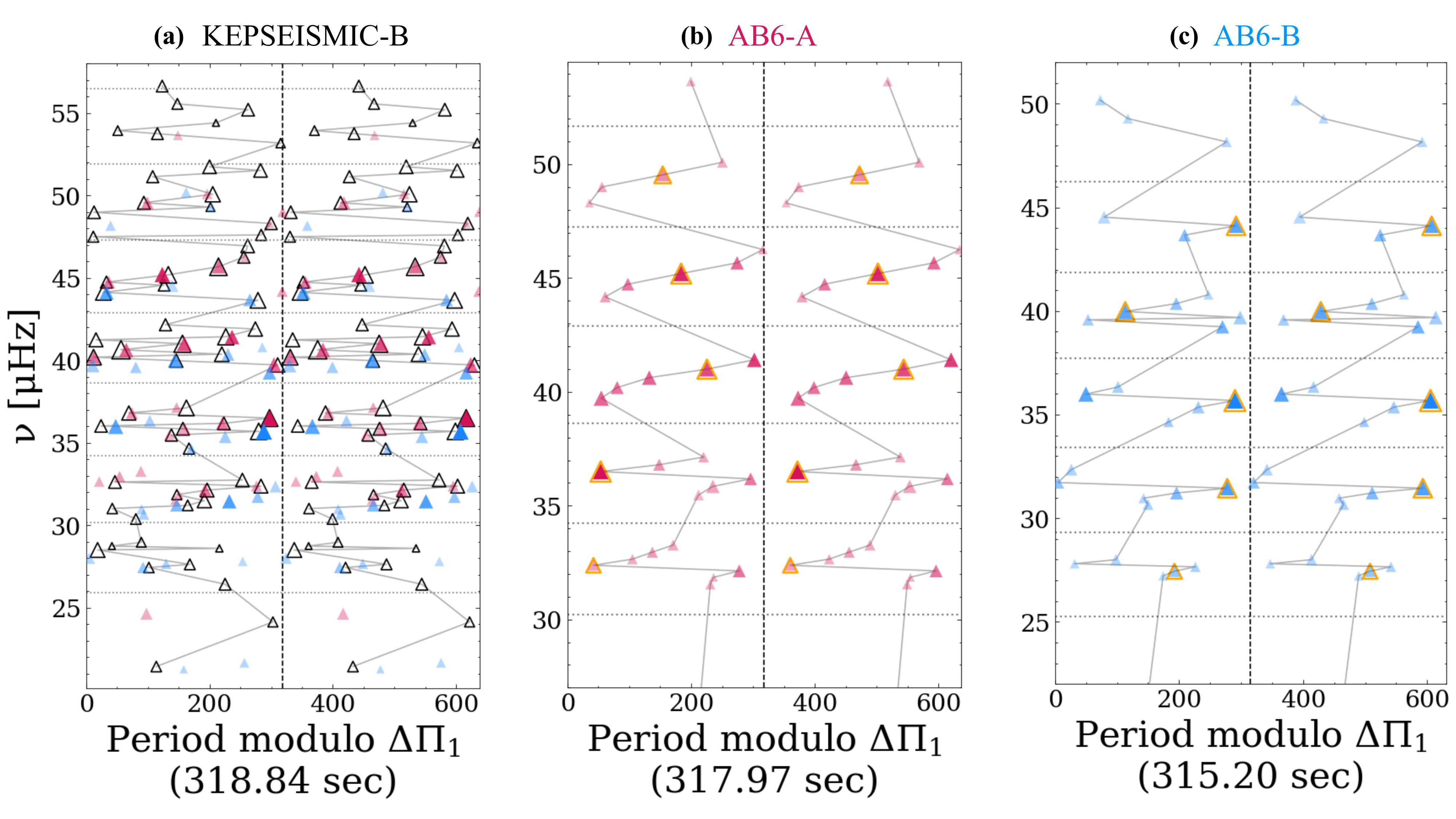}
    \caption{Similar to Fig. \ref{fig:5098_period}, now for AB6, AB6-A and AB6-B. Panels (a) - (c) show period {\'e}chelle diagrams of $\mathrm{\ell=1}$ modes from the PDSs in Fig. \ref{fig:6430834_PDS}(a), (c), and (e), respectively. In panel (a), $\mathrm{\ell=1}$ modes from Fig. \ref{fig:6430834_PDS}(a) are indicated by black triangles, and $\mathrm{\ell=1}$ modes for AB6-A and AB6-B are overplotted in red and blue, respectively.}   
    \label{fig:6430834_period}
\end{figure*}

In the asteroseismic analysis, the oscillation signal in the PDS provides the primary observable for inferring stellar properties. If we analyze a seismically unresolved AB candidate as a single star, the combined and complex oscillation patterns can lead to inaccurate measurements of global seismic parameters. We found that $\mathrm{\nu_{max}}$ becomes unreliable when the power excesses of the two stars overlap in frequency. This is consistent with \citet{2019_Sekaran}, where the $\mathrm{\nu_{max}}$ measurements from their synthetic PDSs with overlapping power excesses are highly unreliable. In particular, they found that $\mathrm{\nu_{max}}$ difference $\lesssim 50\text{ }\mathrm{\mu Hz}$ between two pulsating stars showed large scatter (see their Figures 5 and 6).

The $\mathrm{\nu_{max}}$ bias also depends on the relative flux contribution between the two stars in AB candidates. In the case of AB2, the $\mathrm{\nu_{max}}$ difference between AB2-A and AB2-B is 18.9 $\mu$Hz ($\sim$33\%). The KEPSEISMIC light curves of AB2-A and AB2-B have different relative flux contribution of the two stars (see Fig. \ref{fig:2570370_2570384_PDS}). Since the oscillation signals of AB2-A and AB2-B are clearly detectable in Fig. \ref{fig:2570370_2570384_PDS}(i), both stars likely contribute similar amount of flux in the KEPSEISMIC light curve of AB2-B. Consequently, $\mathrm{\nu_{max}}$ for AB2-B is underestimated by $\sim$19\% (see Fig. \ref{fig:result_scatter}(a)). If we use the KEPSEISMIC light curve of AB2-A to infer $\mathrm{\nu_{max}}$ value of AB2-B, the bias is further underestimated by $\sim$31\%. 

When a light curve contains oscillation of both stars, the KEPSEISMIC light curve of ABi-A (i=1-6) is typically dominated by ABi-A, while the KEPSEISMIC light curve of ABi-B is dominated by ABi-B. Accordingly, the parameters inferred for ABi-A show smaller bias when they are derived from the KEPSEISMIC light curve of ABi-A, rather than ABi-B. Nevertheless, AB6 provides an interesting case. The $\mathrm{\nu_{max}}$ values of AB6-A and AB6-B are nearly identical, so the estimated $\mathrm{\nu_{max}}$ from the PDSs of KEPSEISMIC light curves are not significantly biased and less sensitive to the relative flux contribution of the two stars. 

Besides $\mathrm{\nu_{max}}$ measurements, $\mathrm{\Delta\nu}$ is also significantly biased in several cases. For example, $\mathrm{\Delta\nu}$ values of AB2-B are underestimated by about 35\% (see Fig. \ref{fig:result_scatter}(b)), regardless of whether we use the KEPSEISMIC light curve of AB2-A or AB2-B. This bias arises from the complex oscillation patterns and unreliable mode identification as discussed in Sect. \ref{subsec:PDS}. Fig. \ref{fig:2570370_2570384_frequency_echelle} illustrates the impact of these biases on the frequency {\'e}chelle diagrams. Only the left panel in Fig. \ref{fig:2570370_2570384_frequency_echelle} computed with $\mathrm{\Delta\nu=\text{4.50 }\mu Hz}$ does not show clear ridges of the $\mathrm{\ell=0}$ and $\mathrm{\ell=2}$ modes. Although the stellar structural variations can influence the vertical ridge sequence (\citealt{2010_Miglio, 2011_Mosser, 2015_Vrard, 2026_Hekker}), the poor alignment here is mainly driven by an incorrect $\mathrm{\Delta\nu}$ and the seismically unresolved oscillations of two stars. Similarly, the frequency {\'e}chelle diagram of KEPSEISMIC-B in Fig. \ref{fig:5098_pds}(b) shows the distorted ridge of the $\mathrm{\ell=0}$ and $\mathrm{\ell=2}$ modes due to the superposition of the oscillation modes from AB5-A and AB5-B. 

We list the derived asteroseismic and fundamental stellar parameters for all ABs and the individual stars in Appendix \ref{sec:appendix-params_table}. We also include the determined evolutionary stage of each star based on TACO outputs. In the $\mathrm{\epsilon_{p,central}}-\mathrm{\Delta\nu_c}$ plane, AB1-A, AB2-B and AB3-B lie close to the boundary between the RGB (stars on the red-giant branch with an inert helium core and hydrogen-shell burning) and the secondary clump (2RC; stars that ignite helium under non-degenerate conditions without undergoing a helium-flash). Hence, we combined the diagnostic from \citet{2012_Kallinger} with spectroscopic information and classified both AB1-A and AB2-B as RGB stars, while AB3-B as an 2RC star, respectively (see Appendix \ref{sec:appendix-params_table} for details). In contrast, both the APOKASC3 catalog and \citet{2025_Vrard} classify AB2-B as a CHeB star. This discrepancy possibly arises because different studies used different light curves. If seismically unresolved oscillations are not taken into account, the oscillation peaks from the CHeB star AB2-A can lead AB2-B to appear as if it is a CHeB star. For AB4-A, the diagnostics from the $\mathrm{\epsilon_{p,central}}-\mathrm{\Delta\nu_c}$ diagram points to the star being an RGB star, whereas APOKASC3 and the spectroscopic diagnostics classify the star as a CHeB star (see Fig. \ref{fig:appendix_ES}). Since our analysis relies primarily on \citet{2012_Kallinger}, we classify AB4-A as an RGB. The classifications of the remaining stars in our sample are in agreement with previous studies \citep{2019_Elsworth, 2025_Pinsonneault, 2025_Vrard}. 

With the combination of biases in $\mathrm{\nu_{max}}$ and $\mathrm{\Delta\nu}$, the mass and radius of AB2-B inferred from the scaling relations show the largest biases in our sample (see Figs. \ref{fig:result_scatter}(c) and (d)). However, AB candidates that show only small biases in their derived masses and radii do not necessarily have correct values. From Eqs. \ref{eq:mass_corr} and \ref{eq:radius_corr}, an overestimated $\mathrm{\nu_{max}}$ can be canceled out by an overestimated $\mathrm{\Delta\nu}$, and vice versa. For instance, the $\mathrm{\nu_{max}}$ of AB4-B is overestimated by $\sim$30\% and $\mathrm{\Delta\nu}$ by $\sim$18\% if we use KEPSEISMIC light curve of AB4-A. Nevertheless, the corresponding mass and radius show only little biases and lie close to zero in Figs. \ref{fig:result_scatter}(c) and (d). Likewise, even though both $\mathrm{\nu_{max}}$ and $\mathrm{\Delta\nu}$ for AB5-B are underestimated when we use the KEPSEISMIC light curve of AB5-B, the resulting mass and radius do not show significant biases. Hence, the absence of a significant bias in mass and radius does not imply that ABs can be ignored. 

AB5 is a particularly interesting case because it contains a low mass RC star, AB5-B. If we use the KEPSEISMIC light curve of AB5-B to infer the mass of AB5-B, the mass is found to be $0.78\pm\text{0.02 }{M_{\odot}}$, while the mass derived from the PDS using the KBonus light curve is $1.12\pm\text{0.05 }{M_{\odot}}$ (see Table. \ref{tab:stellar_analysis}). Assuming the mass loss on the RGB driven by the radiation and pulsation, the mass of $0.78 \pm 0.02 {M_{\odot}}$ at a metallicity of [M/H] $\approx$ 0.15 dex (value for AB5-B from APOGEE DR 17) for an RC star lies close to the theoretical lower mass limit at the current age of the universe (see Figure 2 of \citealt{2022_Li}). Stars that fall below the theoretical lower mass limit are often interpreted as the outcome of the non-canonical evolution due to the past binary interactions or merger histories \citep{2022_Li,2024_Matteuzzi,2024_Rui,2025_Singh}. In addition to these scenarios, we present evidence that some of the observed low mass red giants (${M \leq \text{0.8 } M_{\odot}}$; see examples in \citealt{2019_Elsworth}) may have inaccurately estimated masses due to seismically unresolved AB candidates, as illustrated by AB5. 

\subsection{Impact of seismically unresolved oscillations on the period spacing ($\Delta\Pi_1$) and coupling factor ($q$)} \label{subsec:result_dipole}

\begin{figure}[!htbp]
    \centering
    \hspace*{-0.05\linewidth}
    \
    \includegraphics[width=1.05\linewidth]{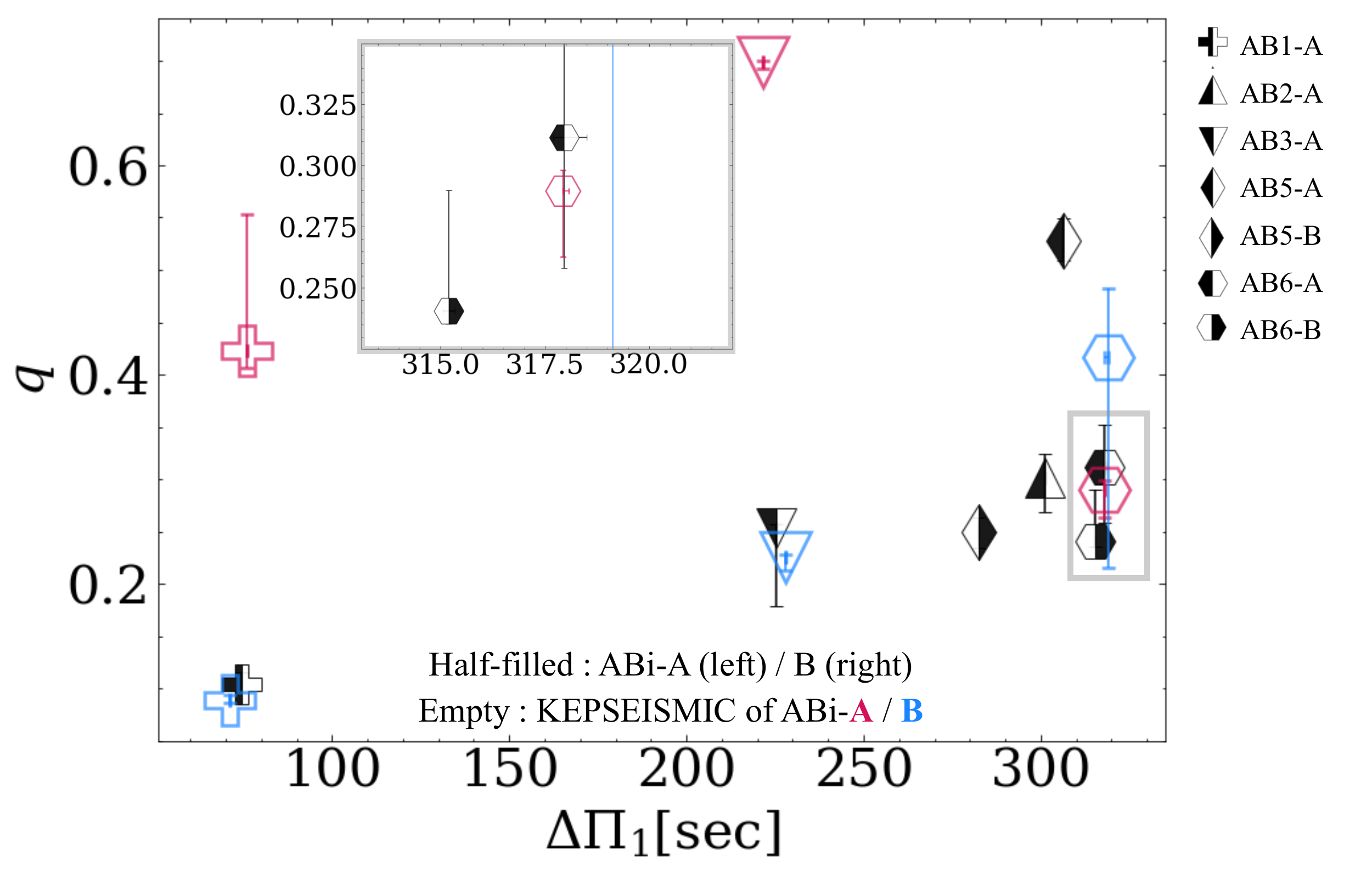}
    \caption{Comparison of the coupling factor $q$ and $\Delta\Pi_1$ between ABs and single stars. Symbols for single stars are the same as in Fig. \ref{fig:result_scatter}. Empty symbols indicate the results from KEPSEISMIC light curves of ABi-A in red and ABi-B in blue (i=1-6). The region enclosed by the gray box is enlarged in the inset.}    
    \label{fig:DPi1_coupling}
\end{figure}

We now examine how seismically unresolved AB candidates bias the $\ell=1$ mixed-modes characteristics. Biases in period spacings ($\mathrm{\Delta\Pi_1}$) and coupling factors ($q$) can lead to incorrect inferences of the internal structures of the stars. We focus on AB5 and AB6, for which we can infer $\mathrm{\Delta\Pi_1}$ and $q$ values for both stars in the AB. 

Fig. \ref{fig:5098_period} presents the stretched period spectra and period {\'e}chelle diagrams of AB5-A and AB5-B. The S/N of AB5-B in panel (b) is more significant compared to AB5-A. This difference arises because the PDS of AB5-A exhibits a dense forest of low S/N $\ell=1$ modes (Fig. \ref{fig:5098_pds}(c)). Thus, for AB5-A, we selected only the $\ell=1$ modes that satisfy $\Delta \text{AIC}\geq10$, indicated by the solid lines in Fig. \ref{fig:5098_period}(c). Since the complex mixed-mode pattern of AB5-A alone already complicates the determination of $\mathrm{\Delta\Pi_1}$ values, it becomes harder to measure $\mathrm{\Delta\Pi_1}$ from the PDS of AB5 (Fig. \ref{fig:5098_pds}(a)). Furthermore, the difference in $\mathrm{\Delta\Pi_1}$ values of AB5-A and AB5-B is about 24 s. As a result, $\mathrm{\Delta\Pi_1}$ from the PDS of AB5 does not converge to a single, unique value. 

In contrast, AB6-A and AB6-B have very similar $\mathrm{\Delta\Pi_1}$ values (differ by only about 3 s; see Fig. \ref{fig:6430834_period}). In this case, we obtained a single $\mathrm{\Delta\Pi_1}$ value for AB6 by using the identified $\ell=1$ modes in Fig. \ref{fig:6430834_PDS}(a). However, the period {\'e}chelle diagram of AB6 in Fig. \ref{fig:6430834_period}(a) shows that this optimized $\mathrm{\Delta\Pi_1}$ value fails to reproduce the $\ell=1$ mode period spacing pattern. A large number of significant $\ell=1$ modes originate from both AB6-A and AB6-B, which create complex $\ell=1$ mode forests. This leads to incorrect determinations of $\mathrm{\Delta\Pi_1}$ and $q$ values when AB6 is treated as a single star. 

In Fig. \ref{fig:DPi1_coupling}, we compare the inferred $q$ and $\mathrm{\Delta\Pi_1}$ between the ABs and their individual stars. The determined values of $\mathrm{\Delta\Pi_1}$ and $q$ along with their uncertainties are in Table. \ref{tab:DPi_q} for each individual star. We show results only when the highest peak in the power spectrum of the stretched-period spectrum reaches S/N > 5. For the cases where a comparison is possible, $\mathrm{\Delta\Pi_1}$ shows minor differences between the AB candidates and their individual stars. However, this does not necessarily indicate the accurate parameters of stars as demonstrated by the case of AB6-A and AB6-B. 

Interestingly, coupling factors are more sensitive to the seismically unresolved AB candidates (see Fig. \ref{fig:DPi1_coupling}). For example, the $q$ value of AB1, inferred from the identified $\ell=1$ modes by using the KEPSEISMIC light curve of AB1-A (Fig. \ref{fig:PDS_2449}(a)), is unusually high ($\sim$0.4) for an RGB star. Since the typical coupling factors of RGB stars decrease with stellar evolution from 0.18 to 0.12 \citep{2017_Mosser_q}, this unusually high $q$ value is unlikely to reflect the intrinsic property of AB1-A. The measurement of the high coupling factor possibly occurs due to the combined oscillation peaks from both AB1-A and AB1-B in the PDS from KEPSEISMIC light curves as shown in Fig. \ref{fig:PDS_2449}(a), which can prevent the reliable $\ell=1$ mode identifications and hence bias the inferred mixed-mode parameters. Furthermore, we cannot robustly determine the mixed-mode parameters of AB1-B that shows low-amplitude dipole modes. Thus, it is important to not use the determined mixed-mode parameters from AB1 for either component.

We found a similar bias in AB3. The PDS yields a significantly overestimated $q$ due to the oscillations from two stars (Fig. \ref{fig:DPi1_coupling}). In particular, the $q$ value of AB3 derived from the PDS in Fig. \ref{fig:PDS_suppressed_3122}(a) is much larger than the value obtained from the PDS of AB3-A in Fig. \ref{fig:PDS_suppressed_3122}(c). Since AB3-B shows low amplitude dipole modes, the results of AB3 from KEPSEISMIC light curve of either AB3-A or AB3-B are not appropriate to analyze AB3-B. 

Lastly, AB6 is the only case for which we can compare the results from the KEPSEISMIC light curves with both AB6-A and AB6-B (see the four hexagon symbols in Fig. \ref{fig:DPi1_coupling}. The derived $q$ of AB6 (from the KEPSEISMIC light curve of AB6-A) is relatively close to that of AB6-A. This is because AB6-A dominates the flux of the light curve. On the contrary, the complex oscillation patterns shown in the PDS of AB6 from the KEPSEISMIC light curve of AB6-B in Fig. \ref{fig:6430834_PDS}(a) leads to notably overestimated $q$. 

\section{Chance alignments and gravitationally bound systems} \label{sec:Discussions}
From the APOKASC3 catalog, we identified six seismically unresolved red-giant asteroseismic binary candidates ($\sim$0.03\% of the total sample in the catalog) that satisfy our selection criteria described in Sect. \ref{subsec:identify_AB}. According to \citet{2025_Mazzi}, red-giant ABs are not expected to interact during their evolution and therefore retain the initial mass ratio close to 1 (e.g., KIC  9246715, an AB composed of two red clump stars with almost similar masses; see \citealt{2016_Rawls}). Additionally, \citet{2025_Mazzi} noted that the ABs consisting of two RC stars are unlikely to survive binary evolution if their orbital separation is similar or smaller than their Roche lobe at the RGB tip (i.e., semi-major axis $\lesssim$ 500 $R_\odot$ or initial orbital periods $\lesssim$ 1000 days). Moreover, \citet{2026_Beck} showed a decrease in the binary fraction as stars ascend the RGB and transition into the CHeB phase, and the systems hosting red giants have longer orbital periods. Recently, \citet{2026_Schimak} found KIC 10841730 as an AB consisting of an RC and an RGB star in a wide binary system. Considering these findings, we applied the wide binary diagnostics from \citet{2025_Espinoza-Rojas}, which is based on the method of \citet{2021_El-badry}. This allows us to investigate whether any of the six AB candidates in our study are gravitationally bound. However, our results indicate that all six AB candidates are most likely chance alignments within the \textit{Kepler} apertures. 

As described in Sect. \ref{subsec:identify_AB}, KIC 3122188 and AB3-B are closely located within the combined TPF of AB3-A and KIC 3122188. AB3-A, KIC 3122188, and AB3-B have RUWE values of about 3.6, 1.7, and 1.0, respectively. Moreover, AB3-A and AB3-B or KIC 3122188 are included in the Washington Double Star (WDS) catalog \citep{2001_Mason}. By checking the astrometric measurements from \textit{Gaia} following the method from \citet{2025_Espinoza-Rojas}, these two stars are potentially in a wide binary system. Moreover, KIC 3122188 was previously reported as a Threshold Crossing Event (TCE; a sequence of transit-like features in the time series that resembles a planetary transit) by \citet{2010_Wu}. The star later classified as a false-positive planetary candidate, potentially due to the contamination from an eclipsing binary. Yet, the nature or the presence of the binary companion is unclear, and it needs further investigation to confirm the binarity.

If AB3-B turns out to be gravitationally bound to KIC 3122188, the system provides an interesting target for studying tidal damping of mixed modes (e.g. \citealt{2013_Ivanov, 2014_Gaulme}), given that AB3-B shows low-amplitude dipole modes. In addition, several studies indicated that oscillations in one component of a gravitationally bounded binary system can be difficult to detect \citep{2016_Rawls, 2025_Mazzi}. This may account for the fact that we only detect oscillation signals from one of the two stars, as shown in Fig. \ref{fig:3122188_TPF}. 

\section{Discussions and Conclusions} \label{sec:Conclusion} 
In this study, we investigated seismically unresolved asteroseismic binary candidates composed of two red-giant stars observed by \textit{Kepler}. We compared the PDSs by using light curves from multiple pipelines, including KEPSEISMIC, PDCSAP, and KBonus, as well as the light curves extracted from custom aperture masks to identify these AB candidates. We identified six seismically unresolved red-giant AB candidates within the APOKASC3 catalog.

We confirmed the predictions from \citet{2025_Choi} by comparing the PDS morphologies of the AB candidates in our study with those of the individual stars. \citet{2025_Choi} concluded that artificial ABs composed of two red clump stars with similar brightness and misaligned mode frequencies make the most complex oscillation patterns in their PDSs, such as in the case of AB6 in our observed cases. If oscillations from two stars overlap in a single PDS, we found cases where the spherical degree of modes are misidentified. 

The presence of three stars with low dipole mode amplitudes (AB1-B in Fig. \ref{fig:PDS_2449}(d), AB3-B in Fig. \ref{fig:PDS_suppressed_3122}(d), and AB4-A in Fig. \ref{fig:PDS_suppressed_404}(c)) among the six AB candidates in our study highlights that seismically unresolved oscillations can strongly bias the observed dipole mode amplitudes and visibilities. Their reduced dipole mode visibilities can be interpreted as signatures of substantial loss of mode energy in the core (e.g., \citealt{2014_Garcia_depressed, 2024_Coppee}), which could be caused by the presence of a strong internal magnetic field (e.g., \citealt{2015_Fuller, 2025_Mueller, 2026_David}, and references therein). Testing such scenarios with observed data requires unbiased measurements of the mode visibilities. Therefore, seismically unresolved AB candidates consisting of suppressed dipole mode stars should be properly accounted for before using their observed visibilities.

If AB candidates are not recognized, the derived asteroseismic and fundamental stellar parameters can be severely biased. In our sample, the biases reach up to $\sim$30\% in $\mathrm{\nu_{max}}$, $\sim$35\% in $\mathrm{\Delta\nu}$, $\sim$150\% in mass, and $\sim$75\% in radius. Moreover, we determined the evolutionary stage of the stars in AB candidates and found that a RGB star (e.g., AB2-B) could be classified as a CHeB star if seismically unresolved oscillations is not taken into account. Furthermore, the period spacings $\mathrm{\Delta\Pi_1}$ and coupling factors $q$ inferred from the PDS that contain oscillation signals of two stars are not trustworthy. Although the $\mathrm{\Delta\Pi_1}$ values obtained for AB candidates and for the individual stars do not exhibit strong biases, they represent inaccurate determinations. Interestingly, the coupling factors show substantial biases and the values inferred from AB candidates in our study tend to be higher than for the individual stars.  

We tested whether any of the identified six seismically unresolved AB candidates are gravitationally bound systems. By applying the method described in \citet{2025_Espinoza-Rojas}, we find that all AB candidates in our study are consistent with chance alignments within the \textit{Kepler} apertures. Additionally, we found a potential wide binary consisting of KIC 3122188 and Gaia DR3 2051928738864291072, for which further observations are required to confirm the nature of the system. 

Our results highlight the importance of identifying and properly accounting for seismically unresolved AB candidates in the asteroseismic analysis. These potential ABs offer explanations not only for the observed stars with complex power density spectra morphologies, but also for red-giant stars with very low masses or very high p- and g- mode coupling factors.

\section*{Data availability}
The light curves extracted by custom aperture masks, and the peakbagging results from TACO will be available in electronic form at the CDS via anonymous ftp to \url{cdsarc.u-strasbg.fr (130.79.128.5)} or via \url{http://cdsweb.u-strasbg.fr/cgi-bin/qcat?J/A+A/}.  

\begin{acknowledgements}
        We thank the anonymous referee for insightful and constructive feedback, which has improved our work. This work is supported by the funding from the ERC Consolidator Grant DipolarSound (grant agreement \# 101000296). In addition, we acknowledge support from the Klaus Tschira Foundation. J.Y.C. acknowledges support from the TOS group at HITS. 
\end{acknowledgements}

\bibliographystyle{aa.bst}
\bibliography{reference}

\begin{appendix} 

\section{Lightcurve extraction from \textit{Kepler} Target pixel files} \label{sec:appendix-TPFs}

\begin{figure}[!htbp]
    \centering
    \includegraphics[width=1\linewidth]{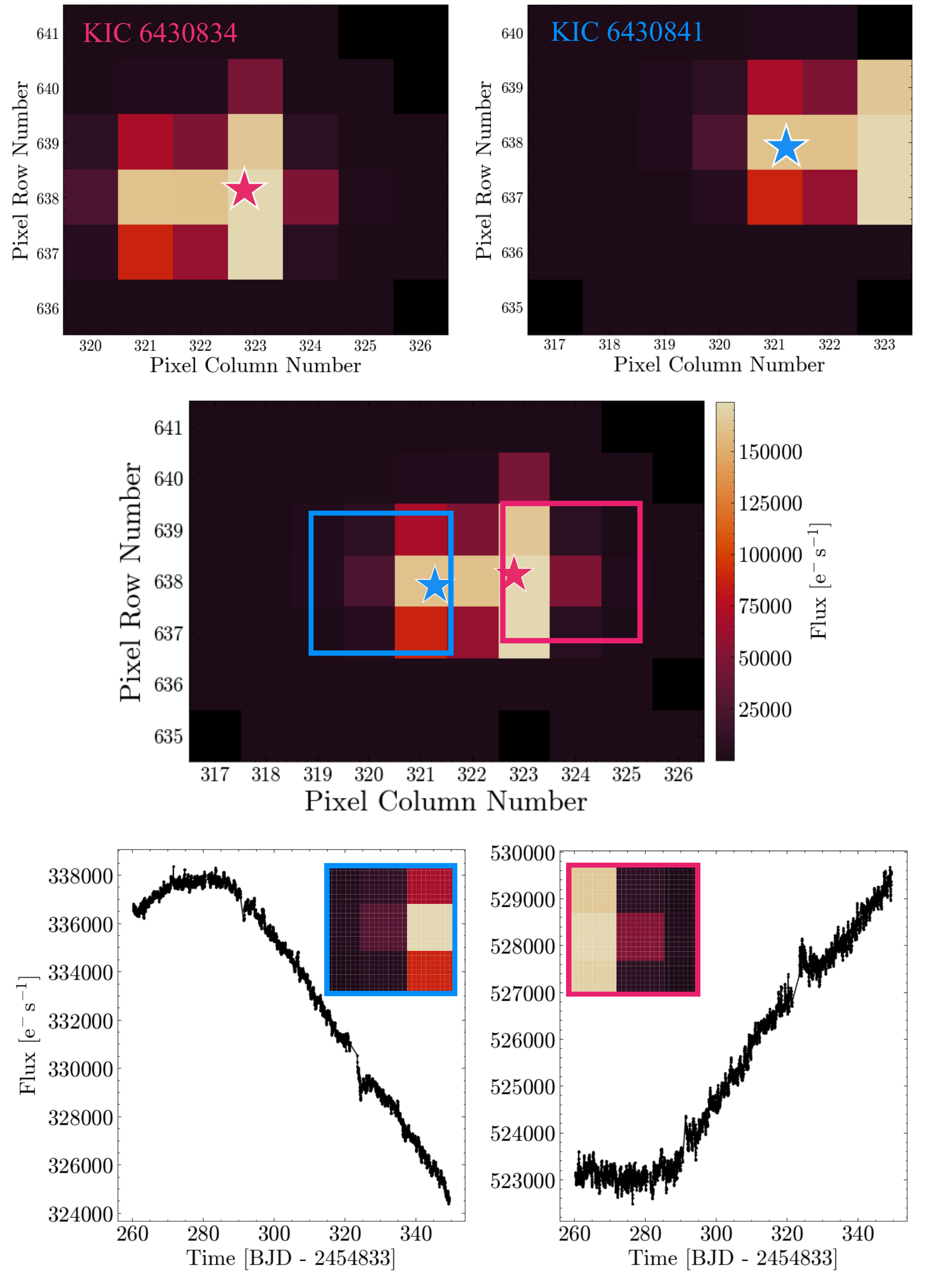}
    \caption{Procedure for creating custom aperture masks and extracting light curves. Top panels: original TPFs corresponding to the first cadence (cadence 7404) of quarter 3 for each star. Middle: combined TPF, with the two custom apertures overplotted in the same colors as corresponding stars. Bottom panels: raw light curves extracted by summing the flux within each aperture. The insets indicate the adopted masks from the combined TPF.}
    \label{fig:appendix_lc_extract}
\end{figure}

As an illustration of the light curve extraction using custom apertures, Fig. \ref{fig:appendix_lc_extract} shows the case of KIC 6430834 and KIC 6430841. The top panels show the original TPF images for each star that we downloaded from the MAST\textcolor{black}{\footnote{\textcolor{black}{https://archive.stsci.edu/missions-and-data/kepler/kepler-bulk-downloads}}}, and their approximate positions are indicated. We then created a combined TPF using the modified \texttt{MOSAIC} pipeline. To place the center of the aperture with higher precision and to define a finely tuned mask, each pixel in this combined TPF was subdivided into a 10 $\times$ 10 sub-pixel grid. This allow us to place the aperture centers on the sub-pixels nearest to the position of the target star. 

In the middle panel of Fig. \ref{fig:appendix_lc_extract}, the two stars are very close to each other on the TPF. To ensure that the two apertures do not share any pixels, the aperture size or location needs adjustments. Since small apertures can exclude too much flux and decrease the S/N of the oscillation signals, we define 26 $\times$ 26 sub-pixel square apertures ($\simeq$3$\times$3 \textit{Kepler} pixel size) and shifted them in opposite directions. For KIC 6430834 (AB6-A) and KIC 6430841 (AB6-B), we shifted the aperture of AB6-A (in red) by 10 sub-pixels towards the right, and that of AB6-B (in blue) by 10 sub-pixels towards the left. The flux within each defined aperture is then summed to produce the individual light curves shown in the bottom panels. Subsequently, we repeated the same procedure for all available quarters. Following \citet{2014_Handberg}, we corrected flux jumps in each quarter flagged by the original \textit{Kepler} data (e.g., reaction wheel desaturation, and telescope attitude tweaks). We used the KASOC (Kepler Asteroseismic Science Operations Center) filter\footnote{\url{https://tasoc.dk/code/_modules/corrections/kasoc_filter/kasoc_filter.html}} as described in \citet{2014_Handberg} to correct the jumps between quarters and stitched all quarters into a single light curve. Then the instrumental drifts, stellar activity, and other sharp features are removed by moving median filters. Finally, we removed 4.5$\sigma$ flux outliers from the light curve.

For the other two stars that we prepared custom light curves for (AB2-B and AB3-B), we also used the same 26 $\times$ 26 sub-pixel aperture size. We shifted the aperture for AB2-B by 10 sub-pixels to the right to reduce flux contamination from AB2-A. In contrast, no shift was required for the aperture of AB3-B. This is because AB3-A and AB3-B are sufficiently separated on the TPF (see Fig. \ref{fig:3122188_TPF}(a)), so that the aperture centered on AB3-B does not significantly overlap with that of AB3-A.

\section{Stars flagged as background sources} \label{appendix:false-positive}
\begin{table}[!htbp]
\centering
\caption{List of stars flagged as background sources in APOKASC3}
\resizebox{0.99\linewidth}{!}{
\begin{tabular}{c c c c}
\hline\hline
 KIC &  \textit{Gaia} ID &   KIC (osc) & \textit{Gaia} ID (osc)  \\
 \midrule
2970580	&2099962209991762048&	2970584&	2099962209991763456\\
$\mathrm{{3103693}^{*}}$ &2099345761927612800&	3103685&	2099345761927612672\\
$\mathrm{{4143910}^{*}}$ &2099656644541165312&  4143913 & 2099656747620593536\\
$\mathrm{{4995049}^{*}}$ &2100701975158974848&  4995064 & 2100701906439499776\\
5557810	&2073839806593689472&		5557808&	2073839806593690496	\\
$\mathrm{{6753396}^{*}}$	&2104744845053412224&		6753393&	2104744840754595840\\
$\mathrm{{9228354}^{*}}$	&2080001156209149312&		9228346&	2080001229227344640\\
9413781	&2080024147173413632&		9413793	&2080024147173414912\\
$\mathrm{{9580167}^{*}}$	&2130359858208650752&	9580171	&2130359858208651008\\
$\mathrm{{9582633}^{*}}$	&2127696841047102464&	9582638	&2127696836746989056\\
9902829	&2080525387035933952&	9902827	&2080525382738321280	\\
$\mathrm{{10081136}^{*}}$&	2128585452601718784&	10081137&	2128585452601719808	\\
10096243& 2085660445637835136	&10096249&	2085659694023032192\\
$\mathrm{{10647686}^{*}}$&	2107679231130471168&	10647684&	2107679231130470784\\
11029161&	2129337101938122752&		11029151&	2129337106238034944	\\
11297580&	2129700666626190208&		11297585&	2129700666629150464	\\
12004941&	2133630218105691008&		12004945&	2133630213812437248	\\
$\mathrm{{12104691}^{*}}$&	2132984770421045888&	12104686&	2132984766121731584\\
\hline \hline
\end{tabular}
}
\tablefoot{Red giants which represent the actual origin of the oscillations are indicated with (osc) next to the column names. Stars also reported in \citet{2019_Hon} is indicated with * next to their KIC IDs. In addition to KIC 3103698 reported by \citet{2019_Hon}, we report KIC 3103693, where both stars resemble the oscillations from KIC 3103685.}
\label{tab:false-positive}
\end{table}

\begin{figure}[!htbp]
    \centering
    \hspace*{-0.06\linewidth}
    \includegraphics[width=0.9\linewidth]{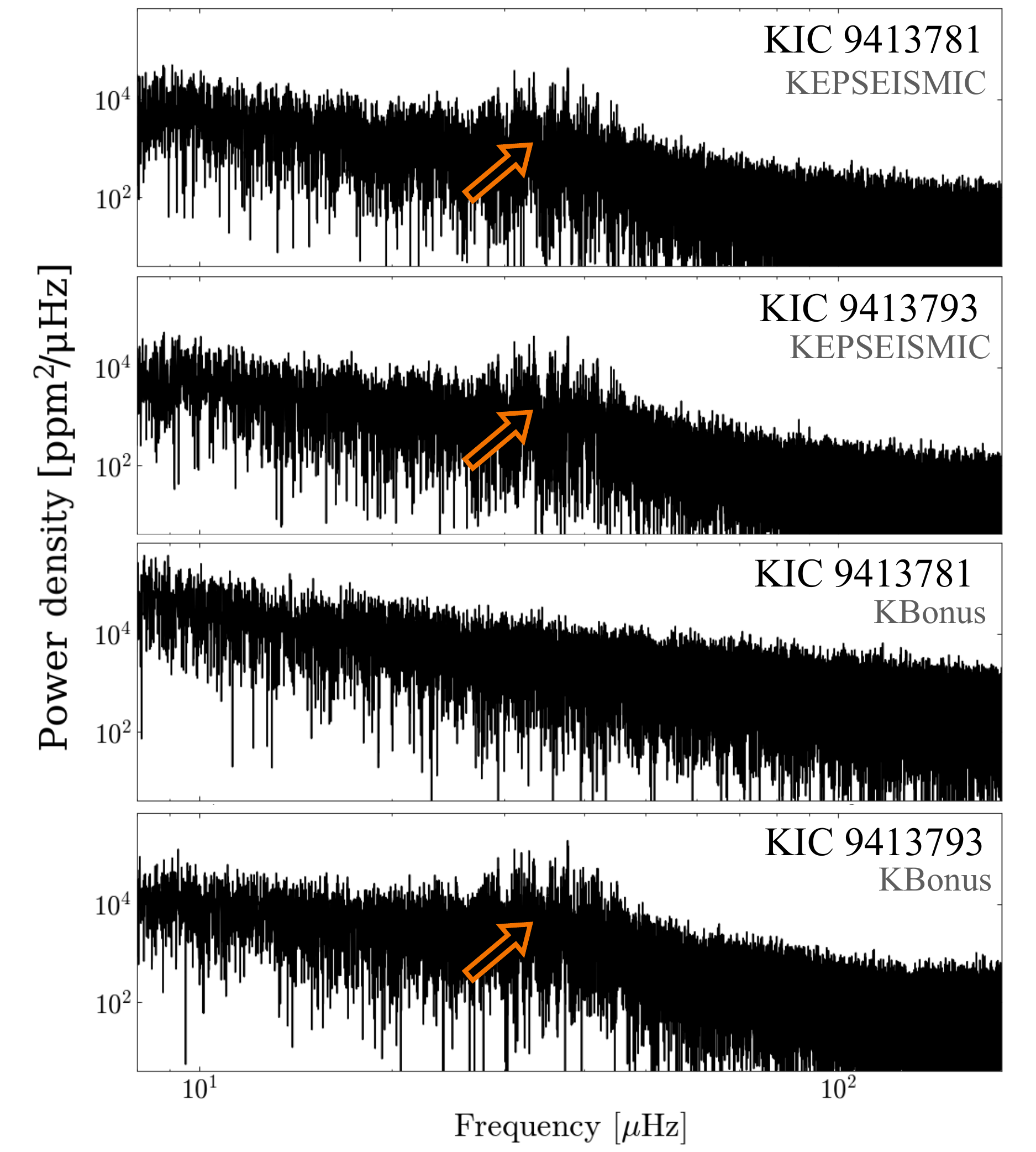}
    \caption{Power density spectra of KIC 9413781 and KIC 9413793, with KIC 9413781 flagged as a background source in APOKASC3 catalog. Top two panels show the PDSs derived from KEPSEISMIC light curves, while the two bottom panels show the PDSs derived from KBonus light curves. Orange arrows indicate the power excesses around $\mathrm{\nu\simeq35\mu Hz}$.}
    \label{fig:appendix_false}
\end{figure}

In Sect. \ref{subsec:candidate_AB}, we selected 40 AB candidates using the reported $\mathrm{\nu_{max}}$ values from the APOKASC3 catalog. Among them, we identified 18 AB candidates that include a star flagged as a background source in APOKASC3. We list 18 such cases in Table \ref{tab:false-positive} and report the true source of the oscillation signals. For the stars listed in the left column of Table \ref{tab:false-positive}, \textit{Kepler} short-cadence light curves are available only for KIC 4995049 and KIC 11297580. We checked these light curves and found no oscillation signals at higher frequencies than the \textit{Kepler} long-cadence Nyquist frequency. Of these 18 stars, APOKASC3 classified 14 stars as dwarfs, while the remaining 4 stars as red giants (KIC 5557810, KIC 9413781, KIC 10096243, and KIC 12004941). KIC 5557810 and KIC 10096243 are also included in \citet{2025_Liagre}, who reported no signals above the \textit{Kepler} long-cadence Nyquist frequency. We additionally examined 4 stars classified as red giants in APOKASC3 using custom light curves (for KIC 5557810 and KIC 10096243) and KBonus light curves (for KIC 9413781 and KIC 12004941), and found no detectable oscillation power excesses. For example in Fig. \ref{fig:appendix_false}, the PDS from the KBonus light curve of KIC 9413781 shows no power excess, which confirms that the oscillation signals detected in the PDS of the KEPSEISMIC light curve for KIC 9413781 are from KIC 9413793. Therefore, the evolutionary stage classification of these 4 stars as red giants is most likely caused by contamination from nearby oscillating red giants. 

\section{AB3 consists of KIC 3122185 and KIC 3122188 or Gaia DR3 2051928738864291072} \label{appendix:AB3}

\begin{figure}[!htbp]
    \centering
    \vspace*{-0.01\linewidth}
    \includegraphics[width=1\linewidth]{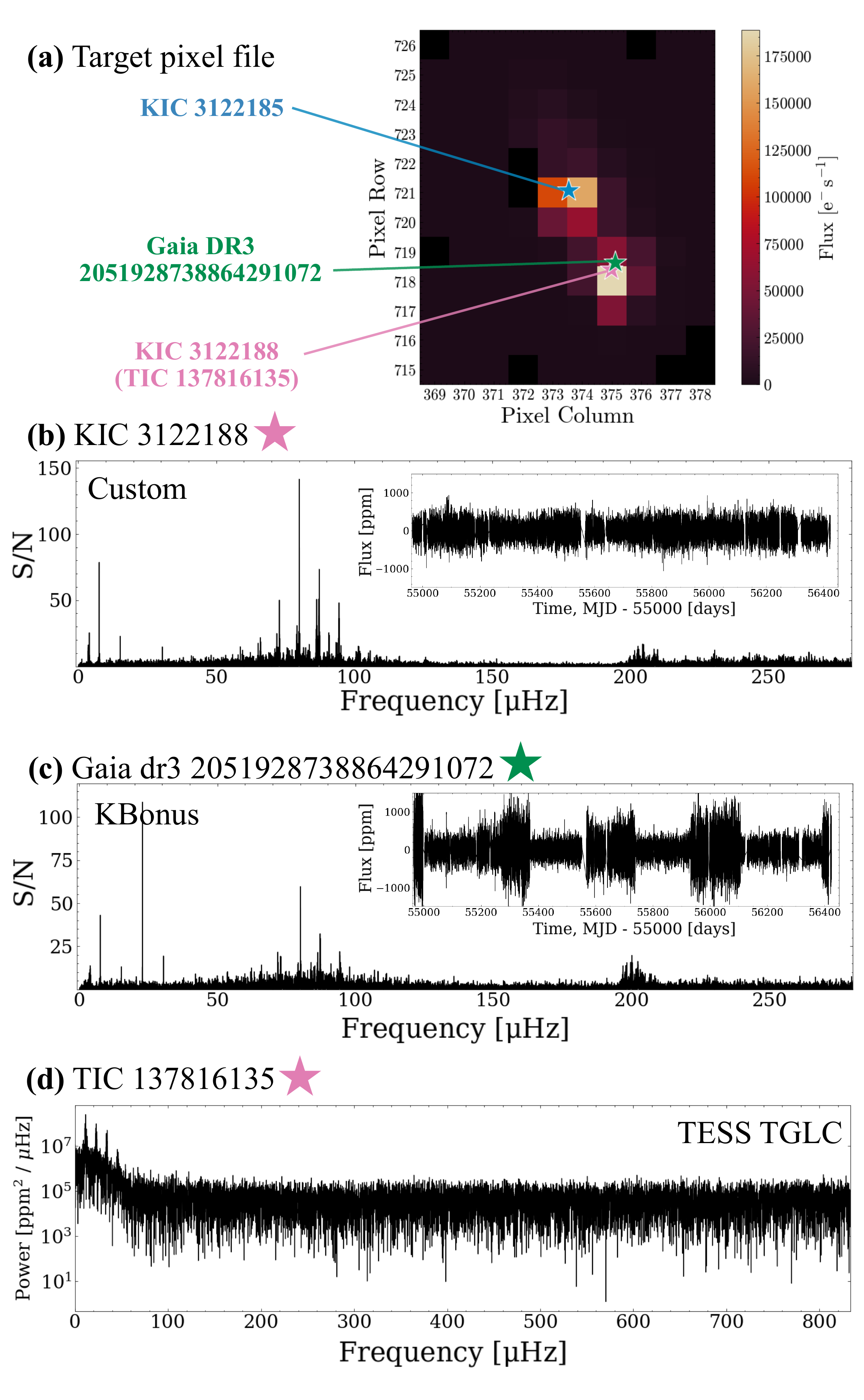}
    \caption{Comparison of the oscillation signals for KIC 3122188 and Gaia DR3 2051928738864291072 (AB3-B). Panel (a): \textit{Kepler} combined TPF of the Quarter 1 of KIC 3122185 (AB3-A) and KIC 3122188. The TPF includes three stars. Panel (b): PDS of KIC 3122188 and the light curve extracted by custom aperture mask is indicated in the inset. Panel (c): Same as panel (b), now for AB3-B with KBonus light curve. Panel (d): PDS of TIC 137816135 derived from TGLC light curve.}
    \label{fig:3122188_TPF}
\end{figure}

Among the 6 identified AB candidates, it is unclear to determine which star is paired with KIC 3122185 (AB3-A) as discussed in Sect. \ref{subsec:identify_AB}. In Fig. \ref{fig:3122188_TPF}, panels (b) and (c) show similar oscillation patterns between $\mathrm{\nu}\sim$ 50 - 100 $\mathrm{\mu Hz}$. The identified oscillation peaks in the panels (b) and (c) of Fig. \ref{fig:3122188_TPF} between $\mathrm{\nu}\sim$ 50 - 100 $\mathrm{\mu Hz}$ appear at nearly identical frequencies, which suggests that one star resembles the oscillation signals from the other. We additionally checked PSF-based TESS-Gaia Light Curves (TGLC; \citealt{2023_Han}) for KIC 3122188 (TIC 137816135, where TIC is the TESS input catalog) with a cadence of 10 minutes in Sector 40, 41 and 54 (see Fig \ref{fig:3122188_TPF}(d)), and did not detect any oscillation signals higher than the Nyquist frequency of the \textit{Kepler} long-cadence data. Therefore, it is not straightforward to attribute the power excess visible at $\mathrm{\nu}\sim$ 50 - 80 $\mathrm{\mu Hz}$ uniquely to either KIC 3122188 or \textit{Gaia} ID 2051928738864291072. Additional constraints from forthcoming \textit{Gaia} data releases and future space missions will be necessary to distinguish the two sources more clearly. 

\section{Asteroseismic and fundamental stellar parameters of 6 AB candidates and the individual stars} \label{sec:appendix-params_table}

\begin{figure}[!htbp]
    \centering
    \includegraphics[width=0.97\linewidth]{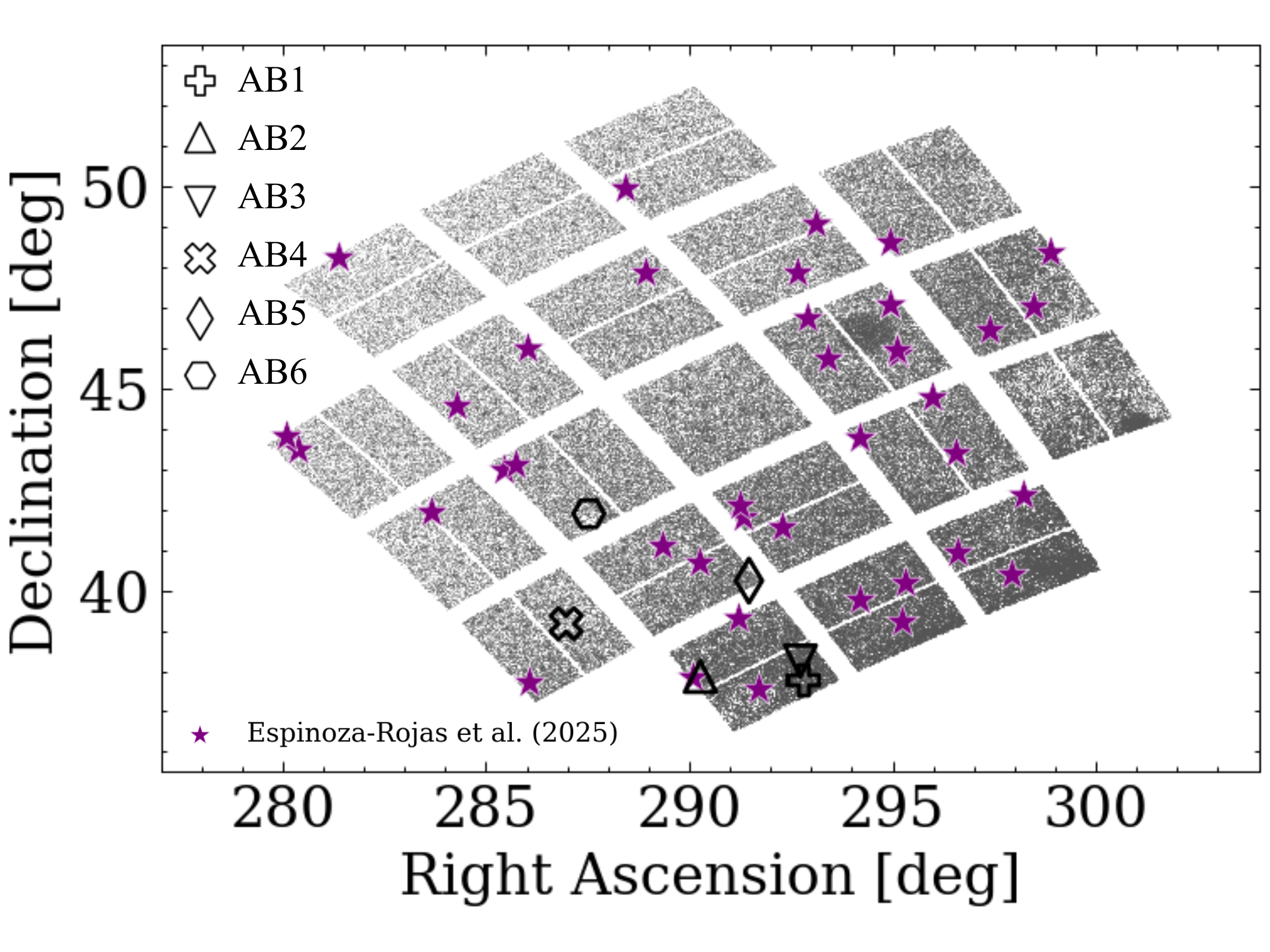}
    \caption{Spatial distribution of \textit{Kepler} targets and KBonus background sources across the \textit{Kepler} field of view, consisting a total of 606,900 stars shown as gray dots. The six seismically unresolved AB candidates analyzed in this work are marked with distinct symbols as described in the legend. Seismically resolved AB candidates from \citet{2025_Espinoza-Rojas} are indicated with purple stars. Note that AB1-AB6 are placed in the dense region.}
    \label{fig:appendix_starmap}
\end{figure}

We summarize the asteroseismic (Tables \ref{tab:seismic_analysis}) and fundamental stellar parameters (Table \ref{tab:stellar_analysis}) of 6 seismically unresolved AB candidates, both derived from the PDSs containing oscillations from two stars and from the PDSs of individual stars. Additionally, we show the spatial distribution of these 6 AB candidates within the \textit{Kepler} field of view in Fig. \ref{fig:appendix_starmap}.

\begin{figure*}[!htbp]
    \centering
    \vspace*{-0.01\linewidth}
    \includegraphics[width=0.89\linewidth]{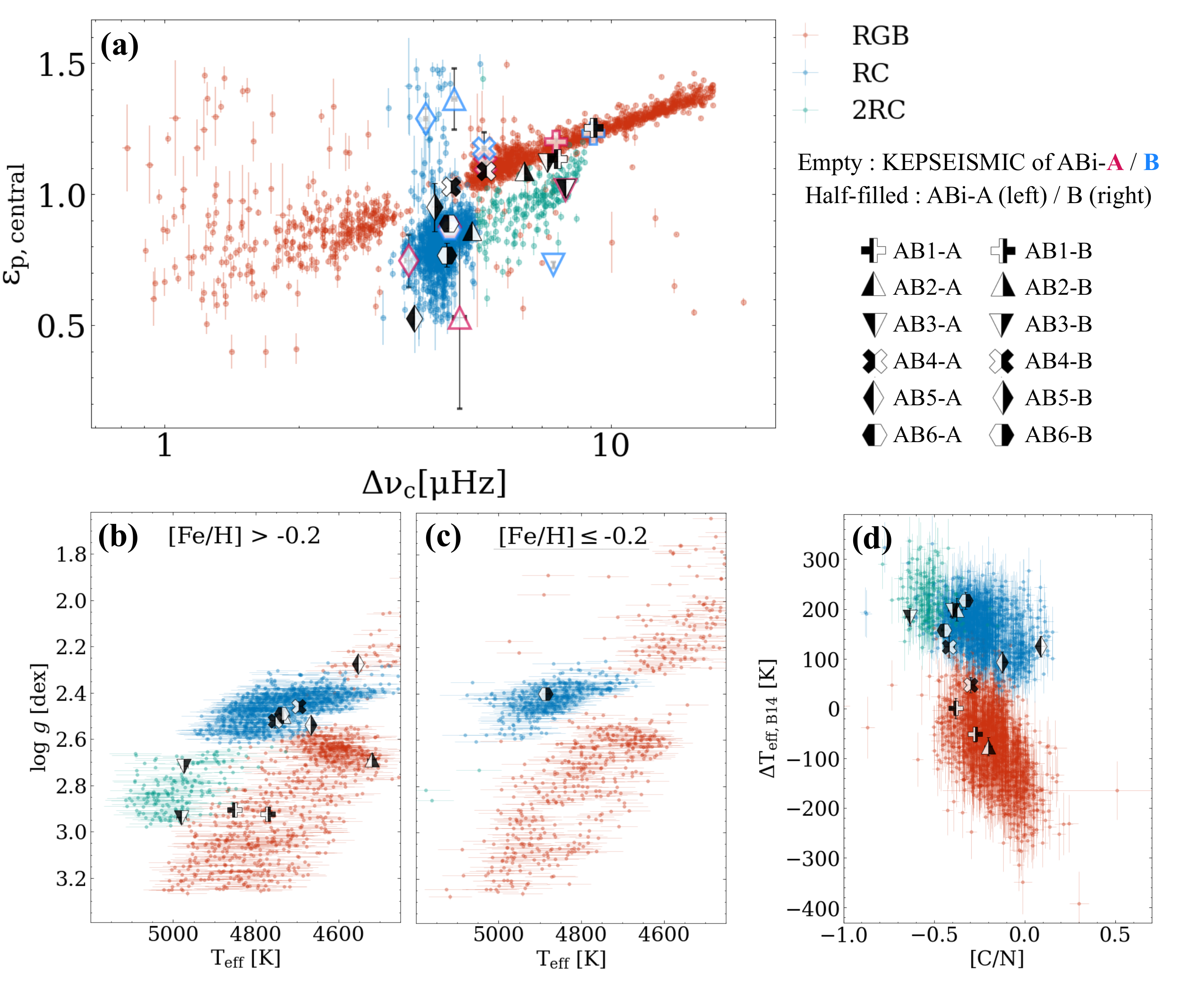}
    \caption{Evolutionary stage determination of individual stars in AB candidates. Symbols are the same as Fig. \ref{fig:result_scatter}, and the circles are from \citet{2019_Elsworth} with their evolutionary stage indicated in the legend. Panel (a): Stars in the $\mathrm{\epsilon_{p,central}}-\mathrm{\Delta\nu_c}$ plane. Results from the PDSs of ABs are also shown with empty symbols. Panels (b) and (c): Distribution of stars in the ${\log g-T_\mathrm{eff}}$ plane for two metallicity ranges divided at $\mathrm{[Fe/H]=-0.2}$, analogous to the Figure 1 of B14. Panel (d): Temperature difference (defined in Appendix \ref{sec:appendix-params_table}) versus the [C/N] ratio of the stars. 
    }
    \label{fig:appendix_ES}
\end{figure*}

Fig. \ref{fig:appendix_ES} illustrates how we infer the evolutionary stage of the stars. We mainly classify stars into RGB, RC and 2RC according to their location in Fig. \ref{fig:appendix_ES}(a). The stars shown as circles in Fig. \ref{fig:appendix_ES} are taken from \citet{2019_Elsworth}, where they consolidated the classification results from different methods. We only selected stars for which all methods in \citet{2019_Elsworth} agree, resulting in a reference sample of 1644 RGB, 1317 RC, and 188 2RC stars. For most stars in our AB candidates, the evolutionary stage is straightforward to determine from Fig. \ref{fig:appendix_ES}(a). However, the classification is ambiguous for AB1-A, AB2-B and AB3-B, as discussed in Sect. \ref{subsec:result_seis_params}. Thus, we used spectroscopic information and further diagnostics described below. 

In Fig. \ref{fig:appendix_ES}(b), both AB1-A and AB2-B can be classified as RGB stars, while AB3-B can be classified as an 2RC star. To confirm this, we considered the log $g$ cut proposed by \citealt{2014_Bovy} (hereafter B14) for selecting CHeB stars (see their equation 2). The upper limit of their equation (2) can be written as
\begin{equation}
0 \le T_{\mathrm{eff}} - \bigg(T_{\mathrm{eff}}^{\mathrm{ref}}([\mathrm{Fe/H}])+\frac{\log g - 2.5}{0.0018}\mathrm{dex \text{ }K^{-1}}\bigg), 
\label{eq:appendix-B14}
\end{equation}
where $T_{\mathrm{eff}}^{\mathrm{ref}}([\mathrm{Fe/H}]) = -382.5\mathrm{K \text{ }dex^{-1}}[\mathrm{Fe/H}] + 4607~\mathrm{K}$. For convenience, we define the temperature difference on the right hand side of Eq. \ref{eq:appendix-B14} as $\Delta T_{\mathrm{eff,B14}}$, such that $\Delta T_{\mathrm{eff,B14}} \geq 0$ indicates CHeB stars. Using $\Delta T_{\mathrm{eff,B14}}$, we plot Fig. \ref{fig:appendix_ES}(d) with [C/N] ratio (see also \citealt{2025_Marasco}). The [C/N] ratio can be used as a proxy for the stellar parameters of red giants \citep{2016_Martig}, and we observe a clear separation between RGB and CHeB (RC and 2RC) stars. As a result, we conclude AB1-A and AB2-B as RGB stars, while AB3-B as an 2RC star. 

\begin{table*}
\caption{Period spacing ($\mathrm{\Delta\Pi_{1}}$) and coupling coefficient ($q$) of seismically unresolved AB candidates}
\centering
\resizebox{0.35\linewidth}{!}{
\begin{tabular}{l c c c}
\toprule   
{KIC} & {Name} & $\mathrm{\Delta\Pi_{1}}$ [s] & $q$\\
\midrule
2449558 & AB1-A & $\mathrm{{74.61^{+0.02}_{-0.03}}}$ & $\mathrm{{0.10^{+0.01}_{-0.01}}}$\\
\midrule
2570370 & AB2-A & $\mathrm{{301.32^{+0.07}_{-0.21}}}$& $\mathrm{{0.30^{+0.02}_{-0.03}}}$\\
\midrule
3122185 & AB3-A & $\mathrm{{225.4^{+0.1}_{-0.2}}}$  & $\mathrm{{0.253^{+0.004}_{-0.074}}}$\\
\midrule
5098143 & AB5-A & $\mathrm{{306.39^{+0.09}_{-0.08}}}$&$\mathrm{{0.53^{+0.02}_{-0.02}}}$\\[2pt]
5098145 & AB5-B & $\mathrm{{282.32^{+0.27}_{-0.04}}}$ & $\mathrm{{0.25^{+0.01}_{-0.02}}}$\\[2pt]
\midrule
6430834 & AB6-A & $\mathrm{{318.0^{+0.5}_{-0.3}}}$& $\mathrm{{0.31^{+0.04}_{-0.05}}}$\\[2pt]
6430841 & AB6-B & $\mathrm{{315.2^{+0.1}_{-0.2}}}$& $\mathrm{{0.24^{+0.05}_{-0.01}}}$\\[2pt]
\bottomrule
\end{tabular}}
\tablefoot{We only report values of the period spacing ($\mathrm{\Delta\Pi_{1}}$) and coupling coefficient $q$ of the individual stars whose highest peak in the power spectrum of the stretched-period spectrum has a S/N > 5. The listed values are obtained from the forward modeling of mixed mode frequencies by BOChaMM pipeline (Sect. \ref{subsec:deltaP_q}).}
\label{tab:DPi_q}
\end{table*}

\begin{table*}
\caption{Global asteroseismic parameters of 6 seismically unresolved AB candidates}
\centering
\resizebox{0.99\linewidth}{!}{
\begin{tabular}{l c c ccc ccc c }
\toprule   
\multirow{2}{*}{KIC} & \multirow{2}{*}{LC} & \multirow{2}{*}{Name} &\multicolumn{3}{c}{$\nu_{\max}$ [$\mu$Hz]} & \multicolumn{3}{c}{$\Delta\nu$ [$\mu$Hz]} & \multirow{2}{*}{$\mathrm{\epsilon_{p,central}}$}\\
\cmidrule(lr){4-6}\cmidrule(lr){7-9}
&  &  &  KEP-A & KEP-B & Single & KEP-A & KEP-B & Single &(single)\\
\midrule
2449558 & KBonus & AB1-A &  \multirow{2}{*}{90.3$\pm$0.5}  & \multirow{2}{*}{103.3$\pm$0.8} & 86.1$\pm$0.4 & \multirow{2}{*}{7.627$\pm$0.002} & \multirow{2}{*}{9.148$\pm$0.003} & 7.555$\pm$0.004 & 1.135$\pm$0.011\\
2449570 & KBonus & AB1-B &  &  & 106.6$\pm$0.5 &  &  & 9.159$\pm$0.005 & 1.256$\pm$0.011\\
\midrule
2570370 & KBonus & AB2-A & \multirow{2}{*}{48.6$\pm$0.6} & \multirow{2}{*}{55.4$\pm$2.8} & 47.8$\pm$0.5 &  \multirow{2}{*}{4.580$\pm$0.151} & \multirow{2}{*}{4.499$\pm$0.024} & 4.823$\pm$0.006 & 0.860$\pm$0.023\\
2570384 & Custom & AB2-B &  &  & 66.7$\pm$0.7 &   &  & 6.433$\pm$0.002 & 1.089$\pm$0.012\\  
\midrule
3122185 & PDCSAP & AB3-A & \multirow{2}{*}{93.3$\pm$1.6} & \multirow{2}{*}{85.7$\pm$0.7} & 99.0$\pm$0.5 & \multirow{2}{*}{7.889$\pm$0.002} & \multirow{2}{*}{7.553$\pm$0.003} & 7.890$\pm$0.003 & 1.045$\pm$0.019\\
... & Custom & AB3-B &   &  & 80.9$\pm$0.6 &   &  & 7.148$\pm$0.004 &  1.120$\pm$0.013\\
\midrule
4042300 & PDCSAP & AB4-A & \multirow{2}{*}{53.6$\pm$2.0} & \multirow{2}{*}{45.4$\pm$0.3} & 55.5$\pm$2.2 &  \multirow{2}{*}{5.227$\pm$0.001} & \multirow{2}{*}{5.097$\pm$0.010} & 5.262$\pm$0.002 & 1.056$\pm$0.019\\
4042308& PDCSAP & AB4-B &  &  & 41.8$\pm$1.1 &   &  & 4.401$\pm$0.002 & 1.026$\pm$0.026\\
\midrule
5098143 & KBonus & AB5-A & \multirow{2}{*}{25.9$\pm$0.3} & \multirow{2}{*}{28.3$\pm$0.3} & 25.2$\pm$0.3 &  \multirow{2}{*}{3.423$\pm$0.004} & \multirow{2}{*}{3.894$\pm$0.003} & 3.490$\pm$0.007 & 0.623$\pm$0.019\\
5098145 & KBonus & AB5-B &  &  & 33.1$\pm$0.5 &   &  & 3.979$\pm$0.019 & 0.951$\pm$0.092\\
\midrule
6430834 & Custom &AB6-A & \multirow{2}{*}{37.6$\pm$0.3} & \multirow{2}{*}{37.1$\pm$0.3} & 38.7$\pm$0.2 &  \multirow{2}{*}{4.324$\pm$0.003} & \multirow{2}{*}{4.302$\pm$0.002} & 4.307$\pm$0.003 & 0.890$\pm$0.002\\
6430841 & Custom &AB6-B &  &  & 35.5$\pm$0.2 &   &  & 4.197$\pm$0.004 & 0.768$\pm$0.045\\
\bottomrule
\end{tabular}}
\tablefoot{The light curves used to create the PDS are specified next to KIC IDs as LC. The "KEP-A/B" columns present the results from the PDSs of KEPSEISMIC light curves for ABi-A/B (i=1-6), respectively. The "Single" columns refer to the results derived from the PDS of an individual star.}
\label{tab:seismic_analysis}
\end{table*}

\begin{table*}
\caption{Fundamental stellar parameters of 6 seismically unresolved AB candidates}
\centering
\resizebox{0.96\linewidth}{!}{
\begin{tabular}{l c cccc ccc c}
\toprule
\multirow{2}{*}{KIC} & \multirow{2}{*}{Name} & \multicolumn{4}{c}{Mass [${M_\odot}$]} & \multicolumn{3}{c}{Radius [${R_\odot}$]} & \multirow{2}{*}{ES}\\
\cmidrule(lr){3-6}\cmidrule(lr){7-9}
 & &   KEP-A & KEP-B & Single & Ratio & KEP-A & KEP-B & Single  &  \\
\midrule
2449558 & AB1-A &  \multirow{2}{*}{1.66 ± 0.03} & \multirow{2}{*}{1.17 ± 0.03} & 1.49 ± 0.02 & \multirow{2}{*}{0.85± 0.02} &  \multirow{2}{*}{7.97 ± 0.05} & \multirow{2}{*}{6.28 ± 0.05} & 7.74 ± 0.04 &  RGB\\
2449570 & AB1-B &  &  & 1.27 ± 0.02 &  &  &  & 6.45 ± 0.03 & RGB\\
\midrule
2570370 & AB2-A & \multirow{2}{*}{2.09 ± 0.26} & \multirow{2}{*}{2.99 ± 0.57} & 1.62 ± 0.05 & \multirow{2}{*}{0.70± 0.03 } & \multirow{2}{*}{12.25 ± 0.74} & \multirow{2}{*}{13.90 ± 0.98} & 10.88 ± 0.12 &  RC\\
2570384 & AB2-B &  &  & 1.14 ± 0.04 &  &  & &  7.82 ± 0.08 & RGB\\
\midrule
3122185 & AB3-A & \multirow{2}{*}{1.82 ± 0.10} & \multirow{2}{*}{1.56 ± 0.04} & 2.17 ± 0.03 & \multirow{2}{*}{0.81 ± 0.02} & \multirow{2}{*}{8.16 ± 0.15} & \multirow{2}{*}{7.89 ± 0.06} & 8.64 ± 0.05 &  2RC\\
... & AB3-B &  &  & 1.76 ± 0.04 &  &  &  & 8.62 ± 0.07 & 2RC\\
\midrule
4042300 & AB4-A & \multirow{2}{*}{1.49 ± 0.17} & \multirow{2}{*}{0.96 ± 0.02} & 1.62 ± 0.20 & \multirow{2}{*}{0.83 ± 0.12} & \multirow{2}{*}{9.86 ± 0.38} & \multirow{2}{*}{8.61 ± 0.07} & 10.08 ± 0.42  & RGB \\
4042308  & AB4-B &  &  & 1.35 ± 0.11 &  & &  & 10.63 ± 0.30 & RGB\\
\midrule
5098143  & AB5-A & \multirow{2}{*}{0.93 ± 0.03} & \multirow{2}{*}{0.78 ± 0.02} & 0.81 ± 0.03 & \multirow{2}{*}{0.72 ± 0.04} & \multirow{2}{*}{11.32 ± 0.12} & \multirow{2}{*}{9.82 ± 0.07} & 10.68 ± 0.11 &  RC\\
 5098145  & AB5-B &  &  & 1.12 ± 0.05 & &  &  & 10.93 ± 0.20 & RC\\
\midrule
 6430834  & AB6-A & \multirow{2}{*}{1.20 ± 0.03} & \multirow{2}{*}{1.19 ± 0.03} & 1.34 ± 0.02 & \multirow{2}{*}{0.90 ± 0.02} & \multirow{2}{*}{10.58 ± 0.08} & \multirow{2}{*}{10.51 ± 0.09} & 11.00 ± 0.06 &  RC\\
 6430841  & AB6-B &  &  & 1.20 ± 0.02 &  & &  & 10.80 ± 0.07 & RC\\
\bottomrule
\end{tabular}}
\tablefoot{Similar structure as Table \ref{tab:seismic_analysis}, now with the mass, radius, and the evolutionary stage (ES) of red-giant stars. "Ratio" denotes the mass ratio for each AB candidate. It uses the masses listed in the "Single" column.}
\label{tab:stellar_analysis}
\end{table*}

\newpage
\section{Comparison with the predicted AB PDS morphologies from \citet{2025_Choi}} \label{sec:appendix-choi}
\citet{2025_Choi} predicted the PDS morphologies of seismically unresolved ABs by creating artificial asteroseismic binaries (AABs). These AABs were created by using KASOC light curves \citep{2014_Handberg}, and varying the flux ratios between the two stars. In contrast, we used light curves extracted by different apertures that changes the total captured fluxes and the flux contribution of each star. Moreover, \citet{2025_Choi} classified the PDS morphologies of AABs showing oscillation signatures from both stars into aligned, partially aligned, and misaligned. Instead, several PDSs of our observed AB candidates exhibit more than one of these categories across the oscillation envelope. For example, AB1-A shows aligned and partially aligned features (Fig. \ref{fig:PDS_2449}(a)), AB3-A shows aligned, partially aligned, and misaligned features (Fig. \ref{fig:PDS_suppressed_3122}(a)), and AB5-A shows partially aligned and misaligned morphologies (Fig. \ref{fig:5098_pds}(a)). Thus, a single category does not describe the PDS morphology of these observed AB candidates. Lastly, the star pairs in AABs have $\mathrm{\nu_{max}}$ separations below 10\%. In our study, only AB6 satisfies this criterion. Therefore, making a direct comparison between our observed AB candidates and \citet{2025_Choi} is not straightforward.

\end{appendix}      

\end{document}